\documentclass[sigconf]{acmart}
\AtBeginDocument{%
  }

\setcopyright{acmcopyright}
\copyrightyear{2023}
\acmYear{2023}
\acmDOI{3544548.3581349}

\acmConference[CHI '23]{CHI '23: ACM CHI Conference on Human Factors in Computing Systems}{April 23-28, 2023}{Hamburg, Germany}
\acmBooktitle{CHI '23: ACM CHI Conference on Human Factors in Computing Systems, April 23-28, 2023, Hamburg, Germany}
\acmPrice{15.00}
\acmISBN{978-1-4503-XXXX-X/18/06}

\acmSubmissionID{9980}

\newcommand{\revised}[1]{{\color{black}{#1}}}
\newcommand{\camera}[1]{{\color{black}{#1}}}

\usepackage{enumitem}

\usepackage{listings}
\usepackage{xcolor,soul}

\definecolor{background}{HTML}{ffffff}
\definecolor{delim}{RGB}{20,105,176}
\colorlet{numb}{magenta!60!black}
\colorlet{punct}{red!60!black}
\lstdefinelanguage{json}{
    basicstyle=\footnotesize\ttfamily,
    numbers=left,
    numberstyle=\scriptsize,
    stepnumber=1,
    numbersep=0pt,
    showstringspaces=false,
    breaklines=true,
    frame=lines,
    backgroundcolor=\color{background},
    literate=
     *{0}{{{\color{numb}0}}}{1}
      {1}{{{\color{numb}1}}}{1}
      {2}{{{\color{numb}2}}}{1}
      {3}{{{\color{numb}3}}}{1}
      {4}{{{\color{numb}4}}}{1}
      {5}{{{\color{numb}5}}}{1}
      {6}{{{\color{numb}6}}}{1}
      {7}{{{\color{numb}7}}}{1}
      {8}{{{\color{numb}8}}}{1}
      {9}{{{\color{numb}9}}}{1}
      {:}{{{\color{punct}{:}}}}{1}
      {,}{{{\color{punct}{,}}}}{1}
      {\{}{{{\color{delim}{\{}}}}{1}
      {\}}{{{\color{delim}{\}}}}}{1}
      {[}{{{\color{delim}{[}}}}{1}
      {]}{{{\color{delim}{]}}}}{1},
}

\definecolor{background}{HTML}{ffffff}
\definecolor{delim}{RGB}{20,105,176}
\colorlet{numb}{magenta!60!black}
\colorlet{punct}{red!60!black}
\lstdefinelanguage{asp}{
    basicstyle=\footnotesize\ttfamily,
    numbers=none,
    numberstyle=\scriptsize,
    stepnumber=1,
    numbersep=0pt,
    showstringspaces=false,
    breaklines=true,
    frame=lines,
    backgroundcolor=\color{background},
    literate=
     *{0}{{{\color{numb}0}}}{1}
      {1}{{{\color{numb}1}}}{1}
      {2}{{{\color{numb}2}}}{1}
      {3}{{{\color{numb}3}}}{1}
      {4}{{{\color{numb}4}}}{1}
      {5}{{{\color{numb}5}}}{1}
      {6}{{{\color{numb}6}}}{1}
      {7}{{{\color{numb}7}}}{1}
      {8}{{{\color{numb}8}}}{1}
      {9}{{{\color{numb}9}}}{1}
      {:}{{{\color{punct}{:}}}}{1}
      {,}{{{\color{punct}{,}}}}{1}
      {\{}{{{\color{delim}{\{}}}}{1}
      {\}}{{{\color{delim}{\}}}}}{1}
      {[}{{{\color{delim}{[}}}}{1}
      {]}{{{\color{delim}{]}}}}{1},
}

\lstdefinestyle{base}{
  moredelim=**[is][\color{red}]{@}{@},
  moredelim=**[is][\color{teal}]{*}{*},
  moredelim=**[is][\color{olive}]{!}{!},
  moredelim=**[is][\color{magenta}]{~}{~},
}

\usepackage{colortbl}
\usepackage{multirow}
\usepackage{float}
\newcolumntype{L}[1]{>{\raggedright\let\newline\\\arraybackslash\hspace{0pt}}m{#1}}
\newcolumntype{C}[1]{>{\centering\let\newline\\\arraybackslash\hspace{0pt}}m{#1}}
\newcolumntype{R}[1]{>{\raggedleft\let\newline\\\arraybackslash\hspace{0pt}}m{#1}}

\usepackage{fontawesome}

\usepackage{scalerel}

\usepackage{subcaption}

\begin{document}

\title{A Review and Collation of Graphical Perception Knowledge for Visualization Recommendation}
\author{Zehua Zeng}
\affiliation{%
  \institution{University of Maryland, College Park}
  \city{College Park}
  \state{Maryland}
  \country{USA}
}
\orcid{0000-0002-5153-3865}
\email{zhzeng@umd.edu}

\author{Leilani Battle}
\affiliation{%
  \institution{University of Washington, Seattle}
  \city{Seattle}
  \state{Washington}
  \country{USA}
}
\orcid{0000-0003-3870-636X}
\email{leibatt@uw.edu}

\renewcommand{\shortauthors}{Zeng and Battle}

\begin{abstract}
  Selecting appropriate visual encodings is critical to designing effective visualization recommendation systems, yet few findings from \revised{graphical} perception are typically applied within these systems.
We observe two significant limitations in translating \revised{graphical} perception knowledge into actionable visualization recommendation rules/constraints: inconsistent reporting of findings and a lack of shared data across studies.
How can we translate the \revised{graphical} perception literature into a knowledge base for visualization recommendation?
We present a review of 59 papers that study user perception and performance across ten visual analysis tasks.
Through this study, we contribute a JSON dataset that collates existing theoretical and experimental knowledge and summarizes key study outcomes in \revised{graphical} perception.
We illustrate how this dataset can inform automated encoding decisions with three representative visualization recommendation systems.
Based on our findings, we highlight open challenges and opportunities for the community in collating \revised{graphical} perception knowledge for a range of visualization recommendation scenarios.
\end{abstract}




\maketitle

\section{Introduction}

\revised{Certain graphical perception results have had a tremendous influence on the design of visualization recommendation systems. For example, Wongsuphasawat et al.~\cite{Wongsuphasawat2015voyager,Wongsuphasawat2017voyager2} leverage theoretical breakthroughs from Bertin~\cite{Bertin1983semiology} and Mackinlay~\cite{Mackinlay1986automating} in the development of the Voyager system. Similarly, Moritz et al.~\cite{Moritz2018formalizing} leverage empirical findings from Cleveland and McGill~\cite{Cleveland1984graphical} and Kim and Heer~\cite{Kim2018assessing} in the development of the Draco recommendation framework. However, we observe many more graphical perception studies that are never considered in the design of visualization recommendation systems.}
\revised{For example, none of the visualization recommendation systems we observe (e.g., as summarized in current surveys~\cite{Zhu2020survey,Shen2021towards,Zeng2021evaluation,Wang2022survey,Wu2022ai4vis}) use guidelines from more than three graphical perception studies to guide their encoding decisions.}
\revised{Some visualization recommendation systems do not reference any graphical perception studies at all to inform their designs (e.g., \cite{Demiralp2017foresight,Key2012vizdeck,Vartak2015seedb}).} 
\revised{We posit that visualization recommendation algorithms could be further enhanced if they could leverage more findings from the graphical perception literature.}

Furthermore, we observe significant \textit{changes and contradictions in encoding guidelines} as knowledge in \revised{graphical} perception continues to evolve~\cite{Kosara2016empire}.
\revised{For example, Cleveland and McGill treated pie charts as primarily angle encodings~\cite{Cleveland1984graphical}; however, more recent work suggests that pie charts are perceived more as area encodings~\cite{Kosara2019evidence}.}
Thus, if a visualization recommendation system only uses a few \revised{graphical perception} papers to guide \revised{its selection of perceptually effective visualizations},
it runs the risk of making \revised{outdated} decisions, which could lead users to misinterpret the data.
\revised{As a result, we argue that graphical perception is a \emph{necessary} component of designing effective visualization recommendation algorithms.}

\revised{Despite the importance of graphical perception in visualization recommendation, we observe that \emph{no current work} establishes a pipeline for integrating graphical perception studies into the design of visualization recommendation algorithms.}
\revised{For example, we fail to find any papers that translate a large body of graphical perception literature into actionable design guidelines for visualization recommendation algorithms~\cite{Quadri2022survey}.} 
Existing surveys either summarize existing \revised{graphical} perception papers~\cite{Ware2012information} or summarize the behavior of existing visualization recommendation systems~\cite{Zhu2020survey,Shen2021towards,Zeng2021evaluation} and ignore
\revised{how one influences the other.}
\revised{Although existing visualization recommendation frameworks~\cite{Moritz2018formalizing,Wongsuphasawat2016towards,Siddiqui2016effortless} enable modeling visualization design knowledge into developing new recommendation algorithms,} users still need to \textit{translate} existing \revised{graphical} perception guidelines into rules/constraints that these algorithms can understand.

We observe two major barriers to translating \revised{graphical} perception
knowledge into actionable visualization \revised{design} rules and constraints: \textit{inconsistent reporting of findings} and \textit{a lack of shared data} across \revised{graphical} perception studies. To address this problem, \revised{one should ideally review the literature in graphical perception, identify which graphical perception studies are actually relevant to visualization recommendation algorithms, and finally} synthesize findings from \revised{the relevant graphical perception studies} in a format that can be \textit{integrated into visualization recommendation code}.

In this paper, we \revised{survey existing graphical perception studies that compare and rank} visualization designs \revised{by perceptual effectiveness} under ten analysis tasks.
We systematically document the visualization designs studied in each \revised{study} and other factors influencing how \revised{visualization} designs are compared, such as input data characteristics. Then, we summarize study outcomes at three levels---between encodings, within chart types, and between chart types---to synthesize concrete \revised{perception-driven design} rules for generating effective visualization designs for specific data characteristics and analysis tasks.
We illustrate how our results can be used to improve existing visualization recommendation systems with three representative systems as case studies: Foresight~\cite{Demiralp2017foresight}, Voyager~\cite{Wongsuphasawat2015voyager,Wongsuphasawat2017voyager2} and Draco~\cite{Moritz2018formalizing}. \revised{Furthermore, we share code to automatically \emph{translate graphical perception results into their corresponding Draco constraints}.}
Finally, we discuss open challenges towards building a knowledge base in \revised{graphical} perception for visualization recommendation, such as contradictory results and missing visualization design pairings in the literature.

\revised{In summary, we make the following contributions in this paper:
\begin{itemize}[nosep]
    \item We review a broad range of the literature (59 papers) on visualization comparison and develop a schema to record the theoretical and experimental results of the comparisons made. The resulting dataset can be ingested into visualization recommendation algorithms to guide the recommendation process.
    
    \item We summarize the major takeaways from graphical perception papers as concrete design guidelines to help visualization recommendation algorithms and even data analysts select optimal visualization designs.
    
    \item \revised{We illustrate how our guidelines could be used to improve existing visualization recommendation systems and share code to translate findings from 30 graphical perception papers into their corresponding Draco~\cite{Moritz2018formalizing} constraints.}
    
    \item Finally, we suggest potential paths for future research to address observed challenges in graphical perception and visualization comparison. 
\end{itemize}

}

All of our data are available online: \url{https://github.com/Zehua-Zeng/graphical-perception-knowledge}.
\section{Related Work}
\label{sec:related-work}

In this section, we discuss existing works in graphical perception and visualization recommendation systems.

\subsection{Graphical Perception Work}
\label{sec:related-work:principles}

Many works investigate how to design effective visualizations.
Theory works such as Bertin's visual encoding principles~\cite{Bertin1983semiology} and Mackinlay's APT work~\cite{Mackinlay1986automating} have been highly influential in information visualization research.
Cleveland \& McGill~\cite{Cleveland1984graphical} organized the encoding channels put forth by Bertin from least to most effective in terms of quantitative data and validated this ranking in part through \revised{graphical} perception studies.
Mackinlay~\cite{Mackinlay1986automating} later extended the ranking to include ordinal and nominal data in the APT system.
Shneiderman~\cite{Shneiderman1996eye}'s task taxonomy then broadened Mackinlay's work by including data types that were not covered in APT, such as multidimensional data, trees, and networks.
The design principles proposed by Bertin, Cleveland \& McGill, Mackinlay, and Shneiderman inform the structure of our framework, which focuses on organizing comparison among not only different visual encodings but also various visualization types.

Numerous later experiments build on these foundational theoretical works.
For example, the experimental results of Cleveland \& McGill were replicated and validated by Heer \& Bostock~\cite{Heer2010crowdsourcing} through crowdsourcing of \revised{graphical} perception experiments.
Talbot et al.~\cite{Talbot2014four} also designed four follow-up experiments on the perception of bar charts to further explore and explain Cleveland \& McGill's results.
Their main goal was to understand how different bar chart designs impact analysis task performance.
Kim et al.~\cite{Kim2018assessing} discuss ways to evaluate the effectiveness of twelve 3-encoding visualization designs for different low-level tasks and dataset characteristics.
Kosara~\cite{Kosara2019evidence} finds that pie charts may be perceived differently than initially hypothesized by Cleveland and McGill.
Saket et al.~\cite{Saket2018task} evaluate the effectiveness of basic visualization types for a specific set of analysis tasks.

\subsection{Visualization Recommendation Systems}
\label{sec:related-work:systems}

\revised{We provide a summary of visualization recommendation systems here and defer to existing surveys for more details~\cite{Zeng2021evaluation,Zhu2020survey,Wang2022survey,Wu2022ai4vis}.}
Existing visualization recommendation systems can be divided into two main categories according to their strategies to rank visualization designs: rule-based or machine learning-based~\cite{Hu2019vizml, Zeng2021evaluation}.
Rule-based systems utilize either existing theoretical principles in \revised{graphical} perception (e.g., \cite{Wongsuphasawat2015voyager, Wongsuphasawat2017voyager2}) or propose new metrics to rank visualization designs (e.g., \cite{Demiralp2017foresight, Vartak2015seedb, Key2012vizdeck}).
For example, Wongsuphasawat et al.~\cite{Wongsuphasawat2015voyager, Wongsuphasawat2017voyager2} use Mackinlay's principles~\cite{Mackinlay1986automating} to make recommendations, prioritizing recommendations based on the breadth of data covered within the visualizations. 
Vartak et al.~\cite{Vartak2015seedb} use an ``interestingness'' metric based on deviation in the data to identify visualizations of potential interest.
Both Key et al.~\cite{Key2012vizdeck} and Demiralp et al.~\cite{Demiralp2017foresight} apply statistical features of the dataset into their systems for guiding exploratory analysis.

\revised{Machine learning-based systems~\cite{Hu2019vizml,Luo2018deepeye,Moritz2018formalizing,Li2022kg4vis,Li2022diverse} design and train models based on (often large) visualization design corpora.}
For example, Hu et al.~\cite{Hu2019vizml} trained a deep learning model using millions of Plotly visualizations and recommended visualization designs for new datasets using the trained model.
In a similar spirit, Luo et al.~\cite{Luo2018deepeye} implemented a visualization recommendation system by combining deep learning techniques with hand-written rules. Moritz et al.~\cite{Moritz2018formalizing} introduced the Draco system, which enables users to generate relevant visualizations by formulating design requirements as rules passed to a constraint solver.
One of the Draco applications, Draco-Learn, was implemented with a training model which learns effectiveness criteria from two prior empirical studies~\cite{Kim2018assessing, Saket2018task}.
\revised{A more recent work by Li et al.~\cite{Li2022kg4vis} proposed a visualization recommendation algorithm based on a knowledge graph employed to model visualization rules, leveraging the advantages of rule-based and machine-learning-based methods.}

\revised{
\subsection{Limitations of Current Work}

All these graphical perception works can (and probably should) inform the design of visualization recommendation systems, yet their influence is still limited~\cite{Saket2018beyond}. Existing visualization recommendation systems only utilize a limited amount of research work as the design guidelines, which introduces the risk of suggesting ineffective visual encodings.
Rather than assessing graphical perception from a structural and implementation perspective, existing surveys primarily summarize graphical perception research to educate non-specialists~\cite{Ware2012information,Ware2008visual}.
We believe this paper is the first to systematically synthesize the graphical perception literature into actionable data and guidelines for visualization recommendation systems.
Furthermore, our work demonstrates how graphical perception work that has generally been overlooked in visualization recommendation systems can be used to improve their performance. 

}
\section{Methodology}
\label{sec:methodology}

Our goal is to enhance the ability of visualization recommendation systems to
reason intelligently about the \textit{effectiveness}~\cite{Mackinlay1986automating} of various visualization designs across analysis tasks and datasets. 
To achieve this, we first need to understand the space of visualization designs and visual comparisons that are most relevant to visualization recommendation systems.
In this section, we formally define the visualization design space that we focus on in this paper.
Then, we describe our method and rationale for collecting and filtering relevant theory and experiment papers in \revised{graphical} perception.

\subsection{Which Visualization Designs Should Be Compared?}
\label{sec:method:design-space}

First, we need to define the visualization design space in which a single recommendation system (or algorithm) can be effective.
On the one hand, it is impractical to derive a single visualization recommendation system to cover all possible visualizations.
On the other hand, it is equally impractical to expect visualization users to learn a completely different system for every conceivable visualization use case.
We establish the boundaries of the visualization design space, which effectively covers the search space for most visualization recommendation algorithms (e.g., Voyager~\cite{Wongsuphasawat2015voyager,Wongsuphasawat2017voyager2}, Foresight~\cite{Demiralp2017foresight}, DeepEye~\cite{Luo2018deepeye}, Draco~\cite{Moritz2018formalizing} etc.). 
Then, we explain how we specify individual visualization designs within this space, informed by the literature on visualization specification and visualization languages~\cite{Satyanarayan2017vegalite,Wickham2010layered}.

\subsubsection{Establishing Design Space Boundaries}
\label{subsec:design-space:boundaries}

Our boundaries are informed by existing literature on (1) visualization design spaces~\cite{Wongsuphasawat2016towards,Moritz2018formalizing,Mackinlay2007showme}, which formally define the range of visualization designs that could be recommended; and (2) \revised{graphical} perception studies~\cite{Tory2006visualization,Van1999cluster,Lee2006task,Heer2007animated,Satyanarayan2017vegalite}, which can be used to identify a subset of designs that can be fairly compared in terms of user performance.
We summarize our findings as the following constraints on the visualization design space.

\paragraph{B1. Exclude 3D visualizations.}
As found in previous work, users often have difficulty in perceiving information from 3D visualizations~\cite{Tory2006visualization}. Most recommendation algorithms do not include them~\cite{Zeng2021evaluation}. 
Moreover, in many cases, multiple linked 2D views prove to be more effective than a single 3D visualization of the same data~\cite{Van1999cluster}. Thus, we exclude 3D visualizations from our design space.

\paragraph{B2. Exclude network graph visualizations.}
As discussed in previous work~\cite{Lee2006task}, graph analysis tasks are generally considered separate from tabular data in visualization research and should likely be studied separately. 
Moreover, existing visualization recommendation systems mainly focus on generating visualizations for tabular data and generally do not include network graph visualizations (e.g., ~\cite{Wongsuphasawat2015voyager,Wongsuphasawat2017voyager2,Moritz2018formalizing,Demiralp2017foresight,Key2012vizdeck}.)
Thus, we exclude graph visualizations, like trees, treemaps, networks, radar charts, chord diagrams, etc.

\paragraph{B3. Focus on static visualization designs.}
Although animations and transitions can improve a user's perception of an underlying dataset~\cite{Heer2007animated}, many if not most visualizations are still designed without any animations or transitions.
Given a lack of data in the literature evaluating the animation and transition design spaces, we do not include these design elements within our visualization design space.
Similarly, the design space of interactions is still an under-explored area in visualization, and enumeration of this space has only recently become viable~\cite{Satyanarayan2017vegalite}.
In this case, the lack of data and theoretical principles is already evident and does not require an in-depth literature review. As a result, we exclude animations and interactions from our analysis. We plan to revisit this gap in our future work as more data becomes available.

\subsubsection{Specifying Visualization Designs}
\label{subsec:design-space:specifying-visualization-designs}

After establishing the design space boundaries, we then discuss how to specify individual visualization designs to be compared.
\revised{The visualization effectiveness could be impacted by many factors, such as the encoding channels, mark types, and scales used in the visualization, but also the data characteristics of the input dataset, like the cardinality and entropy of each attribute. Inspired by one of the most popular visualization grammars, Vega-Lite~\cite{Satyanarayan2017vegalite}, we use data types, data transformations, encoding channels, mark types, and scales to specify each observed visualization design. However, Vega-Lite does not support data characteristics specification. To address this limitation, we extend Vega-Lite by integrating a new ``data characteristics'' component to support describing the target dataset; the specification structure is based on how dataset characteristics are specified in Draco~\cite{Moritz2018formalizing}. Examples of how to use our specification language are provided in \autoref{lst-covered-design} and our supplemental materials.}

\textbf{Data Types:} quantitative, nominal, or ordinal.

\textbf{Data Characteristics:} cardinality and entropy.

\textbf{Data Transformations:} aggregation or bin.

\textbf{Encoding Channels:} position (X/latitude, Y/longitude), length, angle, area, texture, shape, color saturation, color hue, orientation, column, row.

\textbf{Mark Types:} point, line, area-circle, area-rect, area-arc, area-other, text, geoshape, box-plot.

\textbf{Scales:} linear, log, nominal, or ordinal.

We utilize data types, characteristics, and transformations to describe the data while encoding channels, mark types, and scales to specify the visualization design itself.
There exist more measurements for data characteristics like scagnostics~\cite{Wilkinson2005graph}. 
However, scagnostics are mainly used for one chart type--scatterplot, which is already covered by a recent survey~\cite{Sarikaya2017scatterplots}.
In this paper, we focus on comparing different visualization designs instead of emphasizing one or two specific chart types; thus, we select the three most commonly used measurements for data characteristics--cardinality, entropy, and correlation.

\begin{figure}[H]
\centering
 \includegraphics[width=1.0\columnwidth]{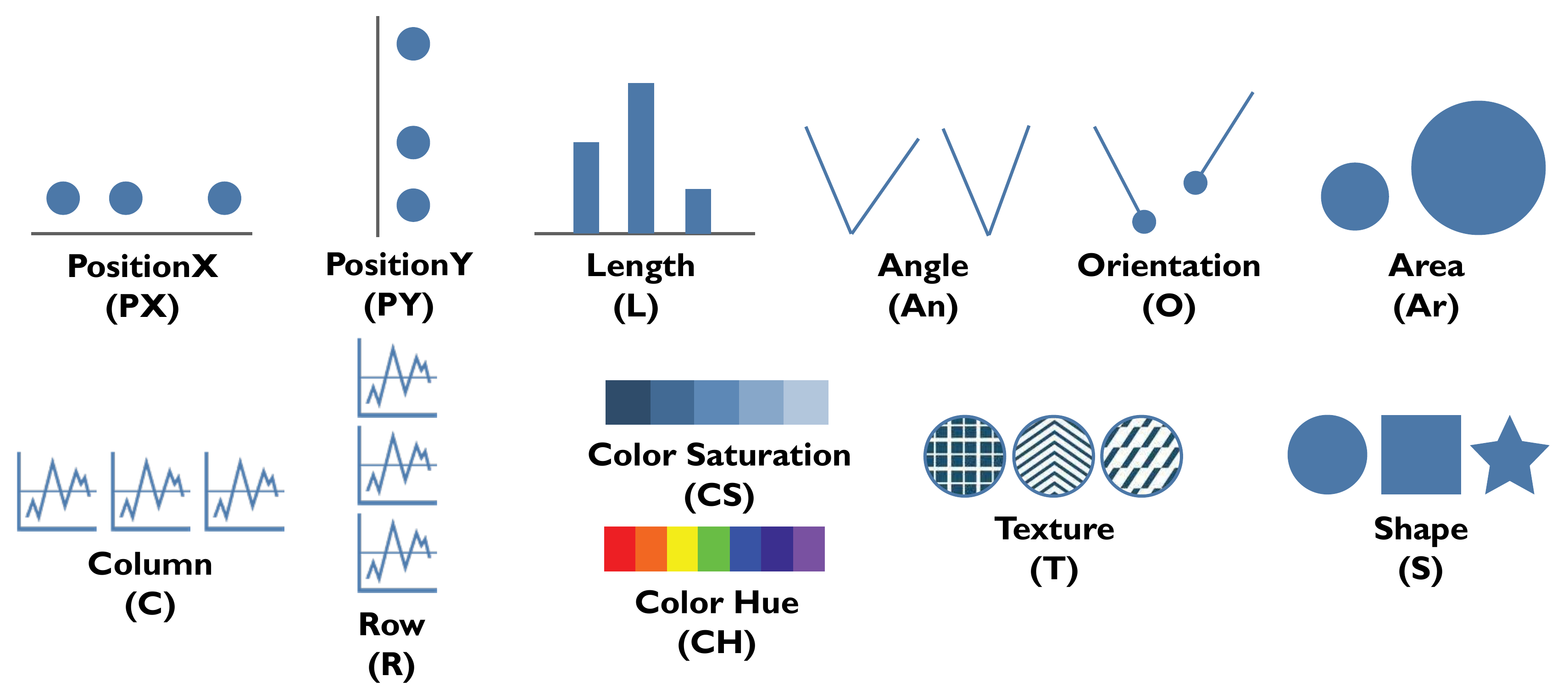}
 \caption{All encoding channels utilized in our design space.}
 \label{fig:encodings}
 \Description{Figure 1 shows all encodings used in our design space, including 12 encodings: PositionX, PositionY, Length, Angle, Orientation, Area, Column, Row, Color Saturation, Color Hue, Texture, and Shape.}
\end{figure}

When selecting encoding channels for our analysis, we start with the encoding channels discussed in the ranking of perceptual tasks proposed by Cleveland \& McGill~\cite{Cleveland1984graphical} and later extended by Mackinlay~\cite{Mackinlay1986automating}. 
We remove the \textsl{connection} and \textsl{containment} channels because they are mainly used for graph visualizations which we exclude from the analysis (\autoref{subsec:design-space:boundaries}).
We also find that \textsl{orientation} has been discussed frequently in the literature (e.g.,~\cite{Chung2016how, Waldner2019comparison}) and is similar to the \textsl{direction} channel proposed by Cleveland \& McGill~\cite{Cleveland1984graphical} and the \textsl{slope} channel mentioned by Mackinlay~\cite{Mackinlay1986automating}; thus we combine them into \textsl{orientation} channel.
We split the \textsl{position} channel into \textsl{positionX} and \textsl{positionY} since there are 2 directions of position in the 2D Cartesian plane, which could impact a user's perception of these values.
We also add \textsl{column} and \textsl{row} encodings for faceting charts, bringing the number of encoding channels to 12 (see~\autoref{fig:encodings}).
To save space, we use an abbreviation to represent each channel.
PX is positionX, PY is positionY, L is length, An is angle, O is orientation, Ar is area, C is column, R is row, CS is color saturation, CH is color hue, T is texture and S is shape, shown in \autoref{fig:encodings}. 

\subsection {Which Papers Should be Included in the Survey?}
\label{sec:method:paper-collection}

To initially find relevant papers for our literature review, we checked all papers in well-known visualization-related conferences and journals (\revised{specifically:} IEEE TVCG, ACM SIGCHI, EuroVIS) in the last ten years, in which we searched for the keywords ``encoding'', ``perception'', ``effectiveness'', ``evaluate'' in the titles, abstracts, and keywords.
We also reviewed the references for each paper found through colleagues or online searches; any relevant papers were also included in our review.
In total, we found 132 candidate papers for our literature review.

We then excluded papers that fall outside the boundaries of the visualization design space described in~\autoref{subsec:design-space:boundaries}.
For example, we excluded papers that only evaluate 3D visualizations, graph visualizations, or animated visualizations.
Given our focus on providing guidelines for visualization recommendation systems, we use the following filters to guide our paper selection process: 

\paragraph{F1. Focus on human perception and task performance.}
An essential facet of visualization recommendation systems is encoding selection, which directly impacts a user's ability to perceive the underlying information~\cite{Kim2017graphscape,Zeng2021evaluation}.
Even if a visualization system suggests certain data attributes to explore, these findings will be inaccessible to the user if the data is presented incorrectly. 
Thus, we focus on results that speak to a user's ability to perceive different visual encodings and differences in user performance across tasks and visualization designs.

\paragraph{F2. Focus on evaluation with standard displays.}
Although some existing work has researched the effect of display size on \revised{graphical} perception or task performance~\cite{Heer2010crowdsourcing, Andrews2010space,Dasgupta2010pargnostics}, and some are building new systems to better support different display sizes~\cite{Brehmer2019visualizing, Badam2018when,Badam2016supporting,Hoffswell2020techniques}, the vast majority of existing visualization evaluations are still conducted in regular displays (e.g., computer screens). 
Thus, we focus on reviewing the literature in visualization evaluation and comparison with standard desktop and laptop displays.

\paragraph{F3. Focus on evaluation with visualizations that can be generated by automatic processes.} 
Recent work~\cite{Cui2019text} combines natural language analysis techniques with visualization synthesis to automatically generate infographics; however, it remains challenging for recommendation systems to understand the semantic meaning of most datasets and then select corresponding encodings or visualizations.
Thus, we exclude papers only evaluating visualizations that usually require a certain amount of manual generation, like visual embellishments~\cite{Skau2015evaluation,Borgo2012empirical,Haroz2015isotype,Skau2017readability}, and semantically color assignments~\cite{Lin2013selecting,Setlur2016linguistic}, etc. 

\paragraph{F4. Compare different visualization designs.}
In order for algorithms to select the most relevant visualization design for a given dataset, they must be able to compare and ultimately rank the effectiveness of different designs~\cite{Zeng2021evaluation}.
To determine which designs should be preferred by these algorithms, we need experimental results that compare different visualization designs or theoretical rules and guidelines to prune irrelevant designs.
To this end, we include any paper in our review that compares the user's ability to effectively perceive and reason about information encoded using different visualization designs (at least one of the six components from~\autoref{subsec:design-space:specifying-visualization-designs} are different).

This filtering step excluded 73 of the 132 candidate papers, leaving 59 papers for our analysis.

\section{Systematically Recording Perceptual Results}
\label{sec:method:schema}

In this section, we present a schema to record extracted visualization rankings.
We use this schema to generate a data record for each of the 59 papers in our literature review, contributing a JSON dataset of \revised{graphical} perception results that can be \textit{imported into visualization recommendation systems}.
Our schema also enables a fine-grained analysis of \revised{how many graphical perception works} exists to inform encoding choices within these systems.
Our schema has four components:

\textbf{Category:} either \textit{theory}, \textit{experiment}, or \textit{hybrid}. \textit{Experiment} papers focus on experiments that provide concrete performance measures for various \revised{graphical} perception scenarios. \textit{Theory} papers present theoretical principles to generalize the findings of empirical work or formal models that can be tested in subsequent work. \textit{Hybrid} papers present a pairing of theoretical hypotheses and experiments conducted to test (at least some of) the proposed hypotheses.

\textbf{Designs:} a list of all the visualization designs tested by each paper in our review, each specified using our visualization space design parameters from~\autoref{subsec:design-space:specifying-visualization-designs}. 

\textbf{Tasks:} a set of visual analysis tasks used in the existing literature for guiding the evaluation of visualization designs.
Previous research has indicated that the effectiveness of visualization depends on the data attributes to be visualized~\cite{Santos2008evaluating} and the task to be performed~\cite{Amar2005lowlevel}; thus, we also include tasks in our schema.
We use prior work~\cite{Amar2005lowlevel,Kim2018assessing} as guidelines and develop a hierarchical task taxonomy.

\camera{\textbf{Results:} \textit{ordered lists} representing the rankings and significant differences reported in the literature for the proposed or tested visualization designs. Results are separated into theoretical rankings and experimental results. 
}

\subsection{Visualization Designs}

\begin{figure}[H]
\centering
 \includegraphics[width=1.0\columnwidth]{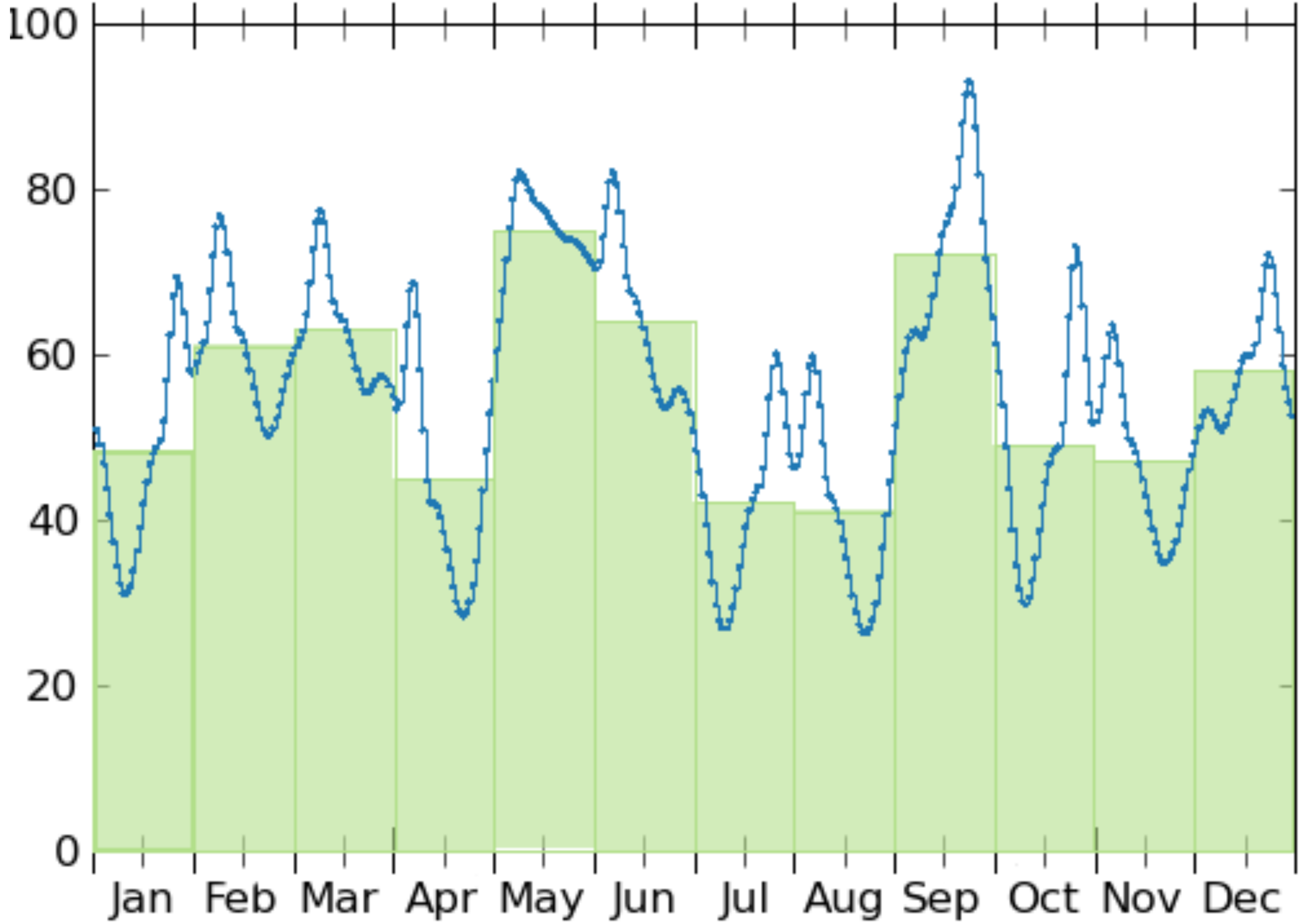}
 \caption{The composite graph studied by Albers et al.~\cite{Albers2014task}. The specification of this chart with our schema is shown in~\autoref{lst-covered-design}.}
 \label{fig:composite-chart}
 \Description{In Listing 1, we use a visualization to demonstrate how our data schema specifies visualization designs. Figure 2 shows the corresponding visualization.}
\end{figure}

Visualization designs are specified by six components as discussed in~\autoref{subsec:design-space:specifying-visualization-designs}. 
We incorporate layers in our schema~\cite{Wickham2010layered} since many designs are only feasible by overlaying multiple visualizations on top of each other.
Each layer is a single visualization design, defined by an encoding set and the mark type.
Each encoding set consists of the data information of the data column visualized by the current encoding and its assigned encoding channel and scale.
~\autoref{lst-covered-design} shows an example of a composite graph, a bar chart (\textsl{line 11-18}) overlaid on a line chart (\textsl{line 3-10}), mentioned in Albers et al.~\cite{Albers2014task} (as shown in~\autoref{fig:composite-chart}).
We assign an ID for each recorded visualization design, where ``E'' means the design is empirically evaluated, and ``T'' means it is theoretically discussed. 
For example, the ID ``E-4'' (\textsl{line 1} in~\autoref{lst-covered-design}) means this visualization design is the fourth empirically tested design in the paper~\cite{Albers2014task}.

When determining the encoding set for a visualization design, we consider all encodings a user or participant perceives within the design rather than the subset of encodings highlighted by a particular experiment or design rule. 
For example, when participants are asked to judge whether test marks are using the same or different colors in scatterplots~\cite{Szafir2017modeling}, 
they perceive three encodings (PX, PY, CH), even if only one encoding (CH) is permuted.


\begin{lstlisting}[language=json,firstnumber=1,caption={Example of a covered visualization design, where a bar chart overlaid on a line chart, as shown in~\autoref{fig:composite-chart}.}, label=lst-covered-design]
"E-4": {
  "layers": [
      {"encodings": [
        { "data-type": "quantitative", "data-charcs": {}, 
          "data-trans": {}, "channel": "positionY",
          "scale": "linear"},
        { "data-type": "ordinal", "data-charcs": {}, 
          "data-trans": {}, "channel": "positionX",
          "scale": "ordinal"}],
        "mark": "line"},
      {"encodings": [
        { "data-type": "quantitative", "data-charcs": {}, 
          "data-trans": {"aggregate": "mean"},
          "channel": "length", "scale": "linear"},
        { "data-type": "ordinal", "data-charcs": {}, 
          "data-trans": {}, "channel": "positionX",
          "scale": "ordinal"}],
        "mark": "area-rect"}]}
\end{lstlisting}

\subsection{Tasks}

\begin{table*}
\caption{A taxonomy of visual analysis tasks based on the task categorizations proposed by Amar et al.~\cite{Amar2005lowlevel} and Kim \& Heer~\cite{Kim2018assessing}.}
 \centering
 \begin{tabular}{C{1em}L{9.5em}L{23.5em}L{17.5em}}
 \toprule
  & \small \textbf{Tasks} &  \small \textbf{Descriptions} &  \small \textbf{Relevant Work}\\
 \hline
 \cellcolor{lightgray!25} &  \small Retrieve Value &  \small Identify values of the specified attributes & \small \cite{Correll2019truncating,Ceja2020truth,Kanjanabose2015multitask,Kim2018assessing,Liu2018somewhere,Mccoleman2021rethinking,Reda2018graphical,Saket2018task,Waldner2019comparison,Sarikaya2017scatterplots,Szafir2016four} \\
 \cellcolor{lightgray!25} &\cellcolor[HTML]{ededed}\small Filter &\cellcolor[HTML]{ededed}\small Find data points satisfying the specified conditions &\cellcolor[HTML]{ededed}\small \cite{Sarikaya2017scatterplots,Golebiowska2020rainbow,Nothelfer2019measures,Nusrat2018evaluating,Saket2018task,Srinivasan2018what,Waldner2019comparison}\\
 \cellcolor{lightgray!25} &  \small Sort &   \small Compare a set of data points by the specified ordinal metric & \small \cite{Correll2019truncating,Lu2021modeling,Sarikaya2017scatterplots,Smart2019measuring,Talbot2014four,Burlinson2018open,Cleveland1984graphical,Heer2010crowdsourcing,Kim2018assessing,Kosara2019impact,Mylavarapu2019ranked,Nothelfer2019measures,Nusrat2018evaluating,Saket2018task,Waldner2019comparison,Zhao2019neighborhood} \\
 \cellcolor{lightgray!25} &\cellcolor[HTML]{ededed}\small Cluster &\cellcolor[HTML]{ededed}\small Detect clusters of similar attribute values &\cellcolor[HTML]{ededed}\small \cite{Demiralp2014learning,Sarikaya2017scatterplots,Szafir2016four,Szafir2017modeling,Wang2018optimizing,Gogolou2018comparing,Kanjanabose2015multitask,Micallef2017towards,Pena2020comparison,Saket2018task}\\
 \multirow{-6}{*}{\rotatebox[origin=c]{90}{\cellcolor{lightgray!25}  \small Value}} &  \small Correlate &  \small Determine/estimate the correlation within the specified attributes & \small \cite{Sarikaya2017scatterplots,Szafir2016four,Aigner2011bertin,Burlinson2018open,Chung2016how,Correll2017regression,Harrison2014ranking,Javed2010graphical,Kanjanabose2015multitask,Kay2015beyond,Li2010judging,Micallef2017towards,Nusrat2018evaluating,Ondov2019face,Pena2019comparison,Reda2018graphical,Saket2018task,Liu2021data,Correll2019truncating} \\
 \cellcolor{darkgray!25} &\cellcolor[HTML]{ededed}\small Aggregate &\cellcolor[HTML]{ededed}\small Compute/compare the aggregate value of the specified attributes &\cellcolor[HTML]{ededed}\small \cite{Gramazio2016colorgorical,Sarikaya2017scatterplots,Szafir2016four,Aigner2011bertin,Albers2014task,Burlinson2018open,Correll2012comparing,Gleicher2013perception,Godau2016perception,Hong2021weighted,Jardine2020perceptual,Javed2010graphical,Kim2018assessing,Mylavarapu2019ranked,Nothelfer2019measures,Nusrat2018evaluating,Ondov2019face,Perin2017assessing,Reda2018graphical,Reda2019evaluating,Saket2018task,Schloss2018mapping,Srinivasan2018what,Xiong2019biased}\\
 \cellcolor{darkgray!25} &  \small Find Extremum &  \small Find data points with an extreme value of the specified attribute & \small \cite{Szafir2016four,Albers2014task,Chung2016how,Golebiowska2020rainbow,Javed2010graphical,Karduni2020bois,Kim2018assessing,Pena2020comparison,Perin2017assessing,Saket2018task,Srinivasan2018what,Waldner2019comparison} \\
 \cellcolor{darkgray!25} &\cellcolor[HTML]{ededed}\small Determine Range &\cellcolor[HTML]{ededed}\small Find the span of values of the specified attributes  &\cellcolor[HTML]{ededed}\small \cite{Albers2014task,Golebiowska2020rainbow,Jardine2020perceptual,Pena2020comparison,Saket2018task,Srinivasan2018what}\\
 \cellcolor{darkgray!25} &  \small Characterize Distribution &  \small Identify the distribution of given attributes & \small \cite{Sarikaya2017scatterplots,Szafir2016four,Albers2014task,Kim2018assessing,Kosara2019impact,Pena2020comparison,Redmond2019visual,Saket2018task} \\
 \multirow{-6}{*}{\rotatebox[origin=c]{90}{\cellcolor{darkgray!25}  \small Summary}} &\cellcolor[HTML]{ededed}\small Find Anomalies &\cellcolor[HTML]{ededed}\small Identify anomalies within the dataset &\cellcolor[HTML]{ededed}\small \cite{Sarikaya2017scatterplots,Szafir2016four,Albers2014task,Burlinson2018open,Gramazio2014relation,Kanjanabose2015multitask,Micallef2017towards,Saket2018task}\\

 \bottomrule
 \end{tabular}
 \label{tab:tasks}
 \end{table*}

Generally, visualizations are designed to facilitate specific visualization tasks or user objectives~\cite{Munzner2009nested, Saket2018task,Rind2016task}, including in visualization recommendation contexts~\cite{Zeng2021evaluation}.
As a result, \revised{graphical} perception experiments gauge performance under a specific subset of tasks.
Furthermore, we and others~\cite{Rind2016task} observe that the literature is not always consistent in how visual analysis tasks are defined.
Thus, we developed a standardized taxonomy to categorize the tasks observed across our selected papers. 
We use the low-level analysis tasks proposed by Amar et al.~\cite{Amar2005lowlevel} as an initial starting point for the taxonomy. 
We use the categorization of Kim \& Heer to divide these low-level tasks into two groups~\cite{Kim2018assessing}: \texttt{value} and \texttt{summary} tasks.
\texttt{Value} tasks require reading or comparing individual values while \texttt{summary} tasks require identification or comparison of the aggregate properties.
Then, we adjusted the task descriptions in response to observed discrepancies from our literature review.
For example, we extended ``Compute Derived Value'' in Amar et al.'s taxonomy into ``Aggregate'' which includes computing and comparing the aggregate values of the specified attributes.
\autoref{tab:tasks} shows all of the visual analytics tasks we observed in our literature review, as well as their descriptions and the relevant works that mention them.

\camera{\subsection{Results}}

\camera{Results represent a summary of the reported outcomes of a given graphical perception experiment or theory paper that can be used to inform visual encoding recommendations.}
When documenting graphical perception results, we distinguish between experimental and theoretical rankings.
They are grouped by the metric that the graphical perception paper uses to rank the visualization designs. 
\camera{We include all and only the metrics that we observed \textit{directly} from the graphical perception literature.}
In total, we observed six different metrics used in the existing literature to compare visualizations: \textsl{accuracy}, \textsl{bias}, \textsl{JND} (just noticeable difference), \textsl{time} and \textsl{user-preference} for experimental results and \textsl{effectiveness} for theoretical rules. 
\camera{Under each metric, we refer to the graphical perception paper or its available supplemental materials to record how visualization designs are ranked based on their task performance (from the best to the worst). }
For visualizations that perform about the same, we group them into a sub-set.
While the \textit{rank} list only reflects the ranking among visualization designs, it does not show whether there is a significant difference between the two visualization designs in terms of each metric.
\camera{Thus, we also record the \textit{significance} results which store a list of visualization pairs where the first entry performs significantly better than the second one. For experimental results, we also record the statistical test results (if reported in the literature), including the statistical method, threshold, and effect size.}

\begin{lstlisting}[language=json,firstnumber=1,caption={Example of the result of a hybrid paper~\cite{Cleveland1984graphical}.}, label=lst-results]
"Results": {
  "Experimental": {
    "accuracy": {
      "sort-1": {
        "rank": ["E-1","E-2","E-3","E-4","E-5"],
        "significance": {
          "pairs":[["E-1","E-3"],["E-1","E-4"], 
          ["E-1","E-5"],["E-2","E-4"],["E-2","E-5"], 
          ["E-3","E-4"],["E-3","E-5"]],
          "significance-method": "bootstrapping", 
          "significance-threshold": 95%,
          "effect-size-method": "none", 
          "effect-size-threshold": null}},
      "sort-2": {
        "rank": ["E-1", "E-6"],
        "significance": {
          "pairs": [["E-1", "E-6"]],
          "significance-method": "bootstrapping", 
          "significance-threshold": 95%,
          "effect-size-method": "none", 
          "effect-size-threshold": null}}}},
  "Theoretical": {
    "effectiveness": {
      "overall": {
        "ranking": ["T-1","T-2",["T-3","T-4","T-5"], 
        "T-6", "T-7",["T-8","T-9"]],
         "significance": {
           "pairs": [["T-1","T-2"], ["T-1","T-3"], 
         ["T-1","T-4"], ..., ["T-7","T-9"]]}}}}}

\end{lstlisting}

We use a hybrid paper~\cite{Cleveland1984graphical} as an example (shown in \autoref{lst-results}).
If a task is conducted multiple times, we add an index after the task name to differentiate different task rounds.
From \autoref{lst-results}, we can see that two \texttt{sort} tasks were conducted (\textsl{line 4, 14}).
\camera{In the first experiment (\textsl{line 4-13}), the \textit{rank} list (\textsl{line 5}) shows that visualization ``E-1'' performed the best among five designs. Moreover, the \textit{significance} result (\textsl{line 6-13}) reveals that bootstrapping method was used to determine whether there is a significant difference between two visualizations, and the threshold was set at 95\%. The \textit{pairs} list reveals that there was a significant performance difference between ``E-1'' and some other visualization designs (``E-3'', ``E-4'', and ``E-5'') (\textsl{line 7-9}). However, the \textit{effect size} was not reported in the experiments from this paper (\textsl{line 12-13}).}


\section{Integrating Current Rankings \& Guidelines}
\label{sec:literature-review}

In this section, we review existing \revised{graphical} perception theories and experiments that could be used to guide visualization recommendation systems.
We \emph{synthesize existing performance rankings} across different visualization designs and summarize the impact of data characteristics and tasks on these rankings. We generate tables summarizing our findings, which system designers can use to specify encoding rules for visualization recommendation systems, e.g., as queries~\cite{Wongsuphasawat2016towards} or constraints~\cite{Moritz2018formalizing}.
We have three research goals for this work: (1) summarize how to rank \textit{individual encodings} according to their expressiveness and effectiveness; (2) summarize how to rank \textit{variations on a single chart type} to enhance its design; and (3) summarize rankings for \textit{comparing different chart types}, to identify the best performing visualization designs for specific data characteristics or task types.

\subsection{Encoding Channels}
\label{sec:litrev-encodings}

\begin{table*}
\caption{Literature coverage for the 12 encoding channels. The papers in italics are \textit{theoretical}, the underlined ones are \underline{hybrid}, and the rest are experimental. }
\centering
  \begin{tabular}{C{2em}|L{46em}}
    \toprule
     & \small\textbf{Relevant Work} \\
    \midrule
  \small PX & \small\cite{Aigner2011bertin,Chung2016how,Albers2014task,Correll2012comparing,Correll2017regression,Gleicher2013perception,Godau2016perception,Gogolou2018comparing,Gramazio2014relation,Harrison2014ranking,Heer2010crowdsourcing,Javed2010graphical,Kanjanabose2015multitask,Kay2015beyond,Kim2018assessing,Liu2018somewhere,Nusrat2018evaluating,Ondov2019face,Reda2018graphical,Saket2018task,Correll2019truncating,Lu2021modeling,Smart2019measuring,Szafir2017modeling,Talbot2014four,Burlinson2018open,Golebiowska2020rainbow,Hong2021weighted,Karduni2020bois,Li2010judging,Liu2021data,Mccoleman2021rethinking,Mylavarapu2019ranked,Nothelfer2019measures,Pena2019comparison,Pena2020comparison,Reda2019evaluating,Schloss2018mapping,Srinivasan2018what,Waldner2019comparison,Xiong2019biased,Zhao2019neighborhood}, \textit{\cite{Mackinlay1986automating,Sarikaya2017scatterplots,Szafir2016four}}, \underline{\cite{Cleveland1984graphical,Wang2018optimizing,Micallef2017towards}}\\
   \small PY &\small\cite{Srinivasan2018what,Aigner2011bertin,Albers2014task,Correll2012comparing,Correll2017regression,Gleicher2013perception,Godau2016perception,Gogolou2018comparing,Gramazio2014relation,Harrison2014ranking,Heer2010crowdsourcing,Javed2010graphical,Kanjanabose2015multitask,Kay2015beyond,Kim2018assessing,Nusrat2018evaluating,Ondov2019face,Reda2018graphical,Saket2018task,Correll2019truncating,Lu2021modeling,Smart2019measuring,Szafir2017modeling,Burlinson2018open,Golebiowska2020rainbow,Hong2021weighted,Li2010judging,Liu2021data,Mccoleman2021rethinking,Mylavarapu2019ranked,Nothelfer2019measures,Pena2019comparison,Pena2020comparison,Schloss2018mapping,Xiong2019biased}, \cellcolor[HTML]{ededed}\textit{\cite{Mackinlay1986automating,Sarikaya2017scatterplots,Szafir2016four}}, \underline{\cite{Cleveland1984graphical,Wang2018optimizing,Jardine2020perceptual,Micallef2017towards}}\\
 \small L & \small\cite{Srinivasan2018what,Albers2014task,Godau2016perception,Harrison2014ranking,Heer2010crowdsourcing,Kay2015beyond,Perin2017assessing,Saket2018task,Skau2016arcs,Correll2019truncating,Lu2021modeling,Szafir2017modeling,Talbot2014four,Ceja2020truth,Karduni2020bois,Kosara2019impact,Mccoleman2021rethinking,Mylavarapu2019ranked,Nothelfer2019measures,Pena2019comparison,Redmond2019visual,Waldner2019comparison,Xiong2019biased,Zhao2019neighborhood}, \textit{\cite{Mackinlay1986automating,Szafir2016four}}, \underline{\cite{Cleveland1984graphical,Jardine2020perceptual}}, \\
 \small An & \small\cellcolor[HTML]{ededed}\cite{Harrison2014ranking,Heer2010crowdsourcing,Kay2015beyond,Skau2016arcs,Lu2021modeling,Kosara2019impact,Ondov2019face,Redmond2019visual,Saket2018task}, \textit{\cite{Mackinlay1986automating}}, \underline{\cite{Cleveland1984graphical}}\\
  \small Ar &\small \cite{Chung2016how,Correll2017regression,Gogolou2018comparing,Harrison2014ranking,Heer2010crowdsourcing,Kay2015beyond,Kim2018assessing,Nusrat2018evaluating,Ondov2019face,Perin2017assessing,Saket2018task,Skau2016arcs,Demiralp2014learning,Hong2021weighted,Javed2010graphical,Kosara2019impact,Mccoleman2021rethinking,Mylavarapu2019ranked,Pena2019comparison,Waldner2019comparison}, \textit{\cite{Mackinlay1986automating,Szafir2016four}}, \underline{\cite{Cleveland1984graphical}}\\
   \small O &\small  \cellcolor[HTML]{ededed}\textit{\cite{Mackinlay1986automating,Szafir2016four}}, \underline{\cite{Cleveland1984graphical}}, \cite{Chung2016how,Harrison2014ranking,Heer2010crowdsourcing,Kay2015beyond,Ondov2019face,Skau2016arcs,Mccoleman2021rethinking,Waldner2019comparison}\\
   \small CH & \small \cite{Srinivasan2018what,Aigner2011bertin,Chung2016how,Correll2012comparing,Gleicher2013perception,Gogolou2018comparing,Gramazio2014relation,Harrison2014ranking,Javed2010graphical,Kim2018assessing,Liu2018somewhere,Reda2018graphical,Saket2018task,Skau2016arcs,Bujack2017good,Bujack2018ordering,Demiralp2014learning,Lu2021modeling,Smart2019measuring,Szafir2017modeling,Golebiowska2020rainbow,Kay2015beyond,Kosara2019impact,Reda2019evaluating}, \textit{\cite{Mackinlay1986automating,Szafir2016four}},  \underline{\cite{Wang2018optimizing,Fang2016categorical,Gramazio2016colorgorical}} \\
  \small CS & \small \cellcolor[HTML]{ededed}\cite{Chung2016how,Albers2014task,Gleicher2013perception,Javed2010graphical,Kim2018assessing,Liu2018somewhere,Ondov2019face,Perin2017assessing,Reda2018graphical,Bujack2017good,Bujack2018ordering,Golebiowska2020rainbow,Hong2021weighted,Mccoleman2021rethinking,Pena2019comparison,Pena2020comparison,Reda2019evaluating,Redmond2019visual,Schloss2018mapping}, \textit{\cite{Mackinlay1986automating,Szafir2016four}}, \underline{\cite{Jardine2020perceptual,Cleveland1984graphical}} \\
   \small T & \small \cite{Chung2016how}, \textit{\cite{Mackinlay1986automating}} \\
  \small S & \small \cellcolor[HTML]{ededed}\cite{Chung2016how,Gleicher2013perception,Kanjanabose2015multitask,Demiralp2014learning,Smart2019measuring,Burlinson2018open}, \textit{\cite{Mackinlay1986automating}} \\
  \small C  & \small\cite{Ondov2019face,Pena2019comparison,Pena2020comparison,Waldner2019comparison}, \textit{\cite{Szafir2016four}}, \underline{\cite{Jardine2020perceptual}}\\
   \small R & \small\cellcolor[HTML]{ededed}\cite{Aigner2011bertin,Javed2010graphical,Ondov2019face,Kim2018assessing,Pena2019comparison,Pena2020comparison,Waldner2019comparison}, \textit{\cite{Szafir2016four}}, \underline{\cite{Jardine2020perceptual}}\\
    \bottomrule
  \end{tabular}
\label{tab:encoding-coverage}
\end{table*}

\revised{In this section, we cluster research papers by the encoding channels they cover. We first discuss how well existing literature covers each encoding channel, then summarize the study outcomes showing how effective each encoding channel is in visualizing different data types.}

\subsubsection{Literature Coverage}
As shown in~\autoref{tab:encoding-coverage}, all twelve encoding channels are covered by existing literature.
To analyze encoding coverage, we break down visualization designs into encoding sets.
For example, a paper studying scatterplots covers both the PX and PY encodings. As another example, a paper that studies grouped bar charts covers three encodings: PX, PY, and CH.
We can see that (PX, PY) (49/59, 83.05$\%$), CH (29/59, 49.15$\%$), L (28/59, 47.46$\%$), and Ar (23/59, 38.98$\%$) are the most discussed encodings, while other encodings such as R (9/59, 15.25$\%$), S (7/59, 11.86$\%$), C (6/59, 10.17$\%$) and T (2/59, 3.39$\%$) are less mentioned.

\subsubsection{Study Outcomes}
\label{subsec:encoding-outcomes}

In~\autoref{tab:mackinlay-encodings}, we summarize findings of encoding perception organized by Mackinlay's principles of \textit{expressiveness} and \textit{effectiveness}~\cite{Mackinlay1986automating}.
A visualization design is considered \textit{expressive} if it shows all and only the data the user wants to see and \textit{effective} if a user can accurately interpret the graphical representation.
We cluster research papers regarding data types to learn which encoding works better for a specific data type.

\begin{table}
\caption{Encoding guidelines summarized from existing \textbf{theoretical} literature. \textbf{Q} means quantitative, \textbf{N} means nominal, and \textbf{O} means ordinal data. An encoding channel is recommended (\faCheck), can partially support (\faAsterisk), or should not be used (\faClose) for the corresponding data type.}
\small
\centering
  \begin{tabular}{l|l|l}
    \toprule
    \multicolumn{3}{c}{\textbf{Expressiveness}}\\
    \midrule
    \textbf{Q} & 
    Ar (\faCheck), CS (\faCheck), CH (\faAsterisk), S (\faClose), T (\faClose) & \\
    \textbf{N} &
    CH (\faCheck), T (\faCheck), O (\faCheck), 
    S (\faCheck), Ar (\faClose), CS (\faClose)  & \\
    \textbf{O} & Ar (\faCheck), CS (\faCheck), T (\faCheck), CH (\faAsterisk), S (\faClose)  & \multirow{-3}{*}{\cite{Mackinlay1986automating,Bujack2018ordering,Bujack2017good}}\\
    \midrule
    \multicolumn{3}{c}{\textbf{Effectiveness}} \\
    \midrule
     \textbf{Q} &   PX$=$PY$>$L$=$An$=$O$>$Ar$>$CS & \cite{Cleveland1984graphical} \\
    \midrule
     \textbf{Q} &  PX$=$PY$>$L$>$An$>$O$>$Ar$>$CS$>$CH &\\
     \textbf{N} &  PX$=$PY$>$CH$>$T$>$CS$>$S$>$L$>$An$>$O$>$Ar & \\
     \textbf{O} &  PX$=$PY$>$CS$>$CH$>$T$>$L$>$An$>$O$>$Ar & \multirow{-3}{*}{\cite{Mackinlay1986automating}}\\
    \bottomrule
  \end{tabular}
\label{tab:mackinlay-encodings}
\end{table}

\begin{figure}[H]
\centering
 \includegraphics[width=1.0\columnwidth]{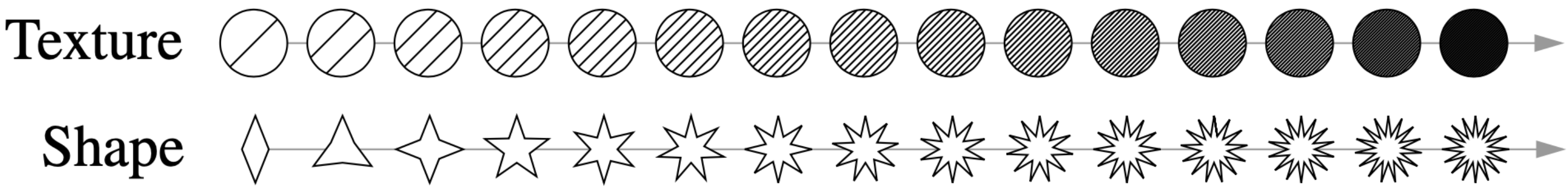}
 \caption{\textit{Texture} and \textit{shape} encodings studied by Chung et al.~\cite{Chung2016how}.}
 \label{fig:chung-encodings}
 \Description{Figure 3 shows the Texture and Shape encodings used in Chung et al.’s experiments.}
\end{figure}

\paragraph{Quantitative.} 
As shown in~\autoref{tab:mackinlay-encodings}, we can see that existing theoretical principles~\cite{Mackinlay1986automating} do not recommend using S and T for quantitative data since they usually cannot be perceived to be ordered. 
However, empirical results from Chung et al.~\cite{Chung2016how} show that S and T \emph{can} be orderable;
in particular, marks with countable differences (e.g., the number of spikes or lines) can be perceived as ordered (see~\autoref{fig:chung-encodings}).

In terms of effectiveness, Cleveland \& McGill~\cite{Cleveland1984graphical} propose a ranking for encodings representing quantitative data, and Mackinlay~\cite{Mackinlay1986automating} extends this ranking to include nominal and ordinal data (see~\autoref{tab:mackinlay-encodings}).
Cleveland \& McGill test \emph{part} of the ranking with follow-up experiments. Their results show that PY encoding outperforms L and An encodings on \texttt{sort} tasks.
Heer and Bostock~\cite{Heer2010crowdsourcing} replicate these experiments but also add Ar encoding in the test and adjust the experiments to make results among tested encodings comparable. 
Their results are similar to Cleveland \& McGill's.
McColeman et al.~\cite{Mccoleman2021rethinking} re-examine these encoding rankings with a different task and find they do not hold. Furthermore, they find that other factors---such as \textit{cardinality---have more influence on task performance than the encoding choice} (see~\autoref{tab:experiment-quant}). 

On the one hand, theoretical works~\cite{Mackinlay1986automating,Bujack2017good,Bujack2018ordering} suggest that the full-color spectrum is not ordered, but part of CH still can be used for quantitative data.
CS, in comparison, is preferable to represent quantitative data.
On the other hand, experiments are conducted to evaluate the human performance of perceiving CS and CH conveying quantitative data with various visual analysis tasks (see~\autoref{tab:color-quant}).
Although Liu \& Heer~\cite{Liu2018somewhere} and Reda et al.~\cite{Reda2018graphical,Reda2019evaluating} arrive at a similar finding which is that participants can discriminate more minor gradient variations with multi-hue colormaps (CH + CS) than with single-hue ones (CS only), they draw different conclusions for the rainbow colormap (\scalerel*{\includegraphics{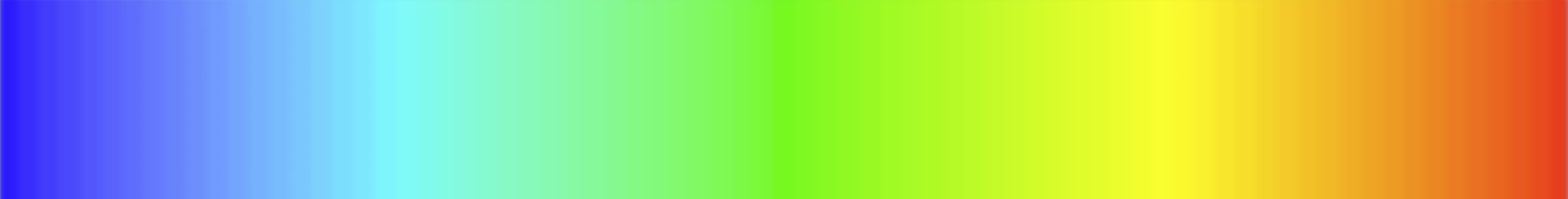}}{B}).
Liu \& Heer~\cite{Liu2018somewhere} suggest that the rainbow colormap is not intuitive and performs the worst for ordering colors and should be jettisoned; however, Reda et al. do not discard the rainbow colormap. They recommend using rainbow or other multi-hue colormaps for \texttt{value} tasks at high spatial frequency.
Schloss et al.~\cite{Schloss2018mapping}, on the other hand, investigate how the different colormaps would be affected by background color.
They find that when colormaps vary less in opacity, human perception is unaffected by the background; however, the role of the background increases when apparent variation in opacity increases.

\begin{table}
\caption{Ranking for encodings representing \textit{quantitative} data from existing \textbf{experimental} literature, group by task type and metric. () differentiates the same encoding but with different mark types. }
\centering
  \begin{tabular}{cccc}
    \toprule
    \small\textbf{Task} & \small\textbf{Metric} & \small\textbf{Rank} & \small\textbf{Ref.}  \\
    \midrule
     &  & \small PY~(bar)$>$L, PY~(bar)$>$An & \small \cite{Cleveland1984graphical}\\
    \multirow{-2}{*}{\small sort} & \multirow{-2}{*}{\small accuracy} & \small PY~(bar)$>$L$>$An$>$Ar & \small \cite{Heer2010crowdsourcing} \\
    \midrule
    \small retrieve value & \small accuracy & \small O$>$Ar$>$CS$>$L$>$PY~(bar)$>$PY~(line) & \\
    \small (2 marks) & \small bias & \small O$>$L$>$PY~(bar)$>$CS$>$Ar$>$PY~(line) & \\
    \cmidrule{1-3}
    \small retrieve value & \small accuracy & \small Ar$>$O$>$CS$>$PY~(bar)$>$L$>$PY~(line) & \\
    \small (4 marks) & \small bias & \small L$>$PY~(bar)$>$O$>$CS$>$Ar$>$PY~(line) & \multirow{-4}{*}{\small \cite{Mccoleman2021rethinking}}\\
    \bottomrule
  \end{tabular}
  \label{tab:experiment-quant}
\end{table}

\begin{table}
  \caption{Ranking for \textbf{colormaps} representing \textbf{quantitative} data, group by task type and metric (only top 3 colormaps are shown). $\gg$ means the left performs better than the right, while $\gtrsim$ means the same order but with some uncertainty.}
  \centering
  \begin{tabular}{cccc}
    \toprule
    \small \textbf{Task} & \small  \textbf{Metric} & \small \textbf{Rank} & \small  \textbf{Ref.}\\
    \midrule
      & \small accuracy  & \small \includegraphics[width=0.065\textwidth]{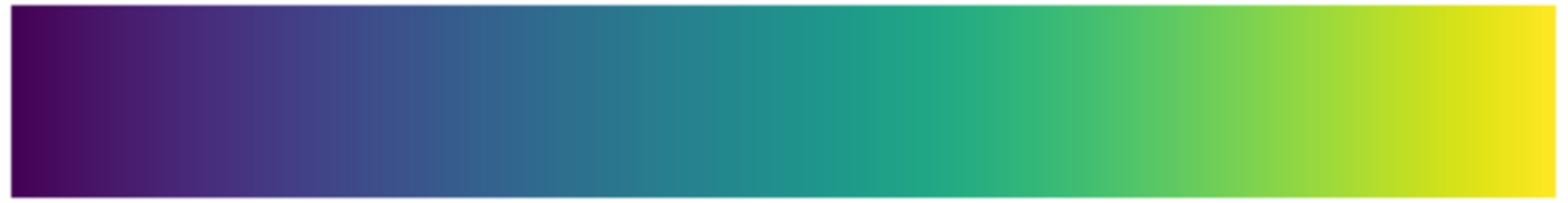} $\gtrsim$ \includegraphics[width=0.065\textwidth]{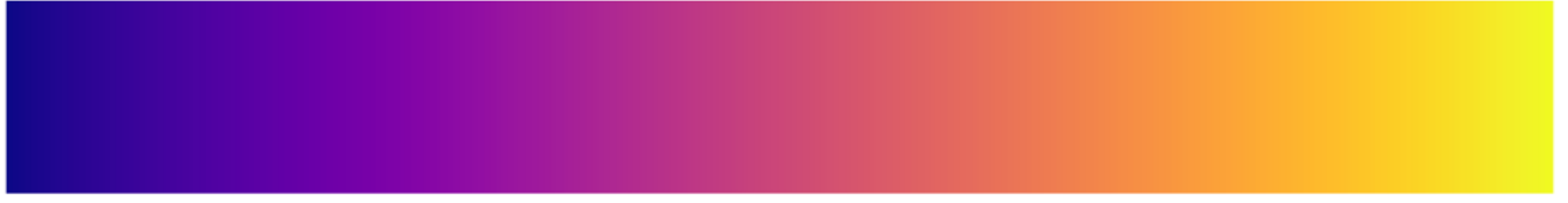} $\gtrsim$ \includegraphics[width=0.065\textwidth]{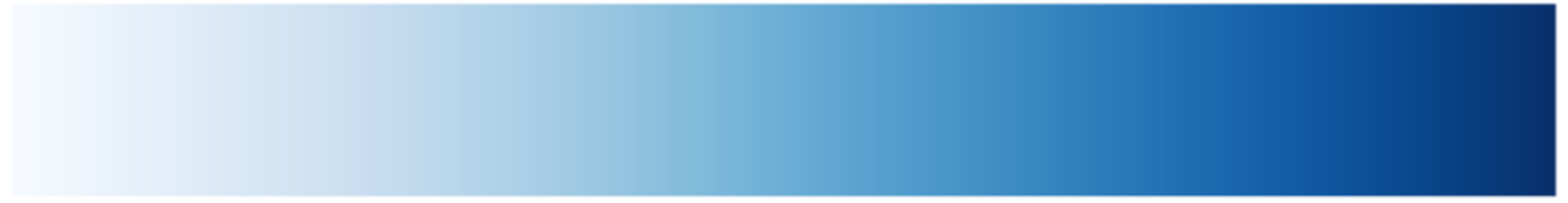} & \\
     \small\multirow{-2}{*}{sort} & \small time  & \small \includegraphics[width=0.065\textwidth]{icons/blues.pdf} $\gtrsim$ \includegraphics[width=0.065\textwidth]{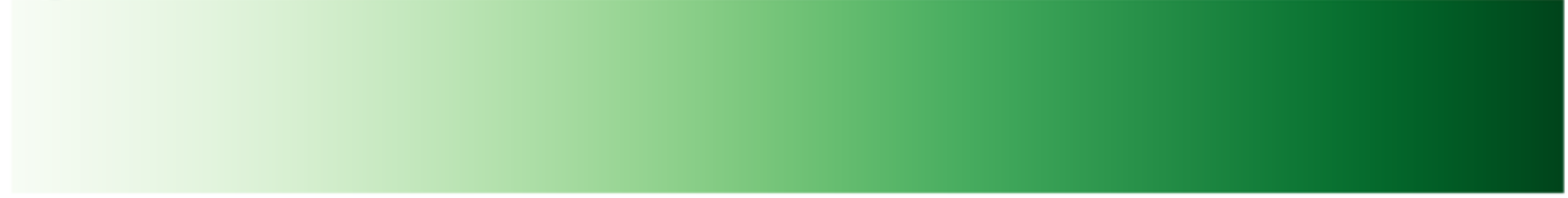} $\gtrsim$ \includegraphics[width=0.065\textwidth]{icons/viridis.pdf} & \multirow{-2}{*}{\small\cite{Liu2018somewhere}} \\
     \midrule
     \small aggregate & \small \small JND & \small \includegraphics[width=0.065\textwidth]{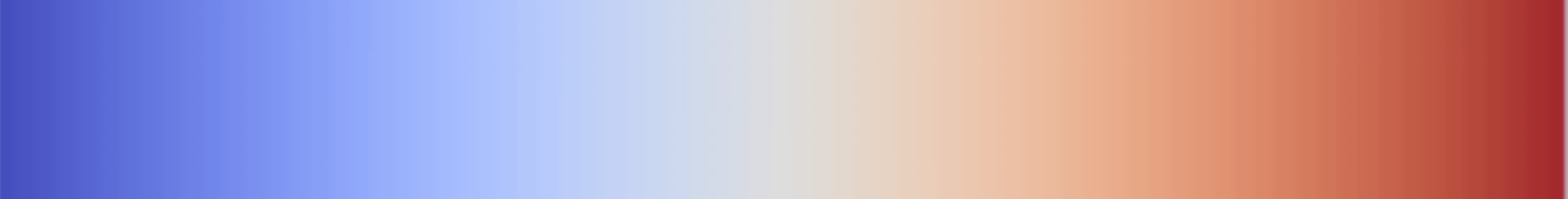} $\gtrsim$ \includegraphics[width=0.065\textwidth]{icons/rainbow.pdf} $\gg$ \includegraphics[width=0.065\textwidth]{icons/viridis.pdf}& \small\cite{Reda2019evaluating} \\
     \midrule
     \small filter &  & \small \includegraphics[width=0.065\textwidth]{icons/rainbow.pdf} $\gg$ \includegraphics[width=0.065\textwidth]{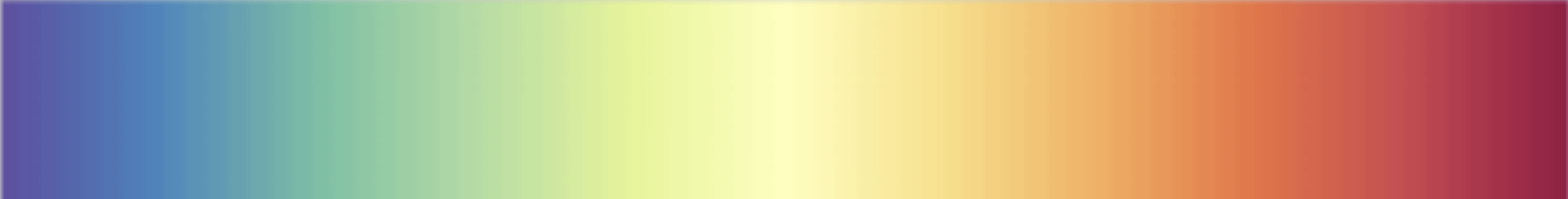} $\gg$ \includegraphics[width=0.065\textwidth]{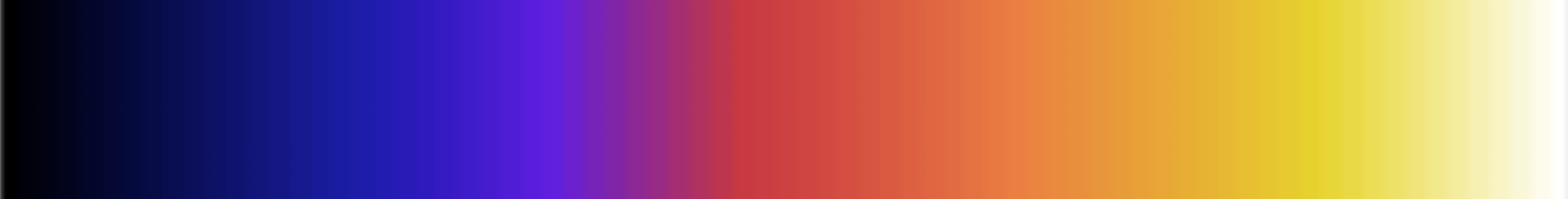}&\\
     \small correlate &  & \small \includegraphics[width=0.065\textwidth]{icons/coolwarm.pdf} $\gg$ \includegraphics[width=0.065\textwidth]{icons/rainbow.pdf} $\gg$ \includegraphics[width=0.065\textwidth]{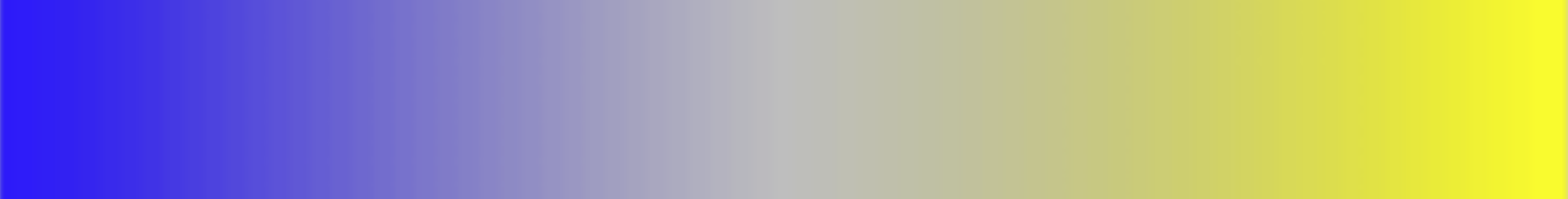}&  \\
     \small cluster & \small \multirow{-3}{*}{accuracy} & \small \includegraphics[width=0.065\textwidth]{icons/coolwarm.pdf} $\gg$ \includegraphics[width=0.065\textwidth]{icons/spectral.pdf} $\gg$ \includegraphics[width=0.065\textwidth]{icons/blueyellow.pdf}& \small\multirow{-3}{*}{\cite{Reda2018graphical}} \\
    \bottomrule
  \end{tabular}
  \label{tab:color-quant}
\end{table}

\begin{figure}[H]
\centering
 \includegraphics[width=1.0\columnwidth]{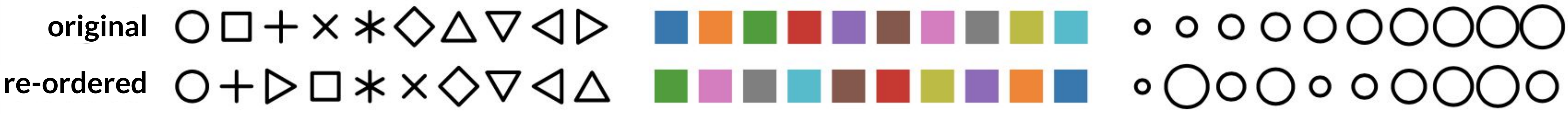}
 \caption{\textit{Shape}, \textit{color hue} and \textit{area} palettes proposed by Demiralp et al.~\cite{Demiralp2014learning}: bottom palettes are re-ordered to maximize perceptual distance.}
 \label{fig:learning-kernels}
 \Description{Figure 4 shows the shape, color hue, and area palettes proposed by Demiralp et al.}
\end{figure}

\paragraph{Nominal.}
Mackinlay's expressiveness rules---the only relevant theory work observed---do not recommend using Ar and CS for nominal data since they would probably be perceived to be ordered.
Empirically-focused papers~\cite{Demiralp2014learning,Smart2019measuring} aim to learn how discriminable different encoding channels are for nominal data.
Demiralp et al.~\cite{Demiralp2014learning} examine the perceptual distance for CH, S, Ar and their combinations.
Based on the experiment results, they propose palettes that can maximize perceptual discriminability for each examined encoding (see~\autoref{fig:learning-kernels}). 
Smart \& Szafir~\cite{Smart2019measuring} measure how using CH, S, and Ar for marks influences data interpretation in multiclass scatterplots.
They find that (1) S affects CH and Ar difference perception more strongly than CH or Ar affects S perception; (2) CH are generally more discriminable with filled shapes than with unfilled ones, and filled shapes (e.g.~\faCircle,~\faSquare~) are perceived as larger than their unfilled counterparts (e.g.~\faCircleThin,~\faSquareO~); and (3) shapes with top or bottom edges (e.g.~\faSquare,~\faSquareO~) are perceived larger than others (e.g.~\faPlus,~\faStarO~).

\paragraph{Ordinal.}
As for expressiveness, existing theory work does not recommend CH to represent ordinal data and recommends CS instead~\cite{Mackinlay1986automating}.
We find one empirical work evaluating CS and CH for conveying ordinal data~\cite{Golebiowska2020rainbow}. They confirm that CS (\scalerel*{\includegraphics{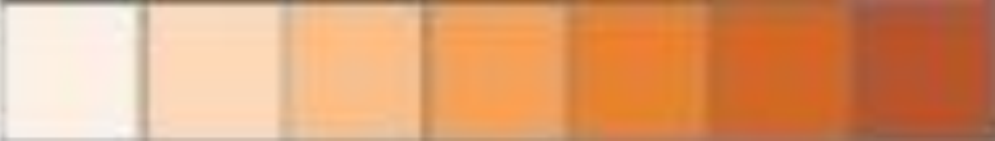}}{B})
performs better than CH (\scalerel*{\includegraphics{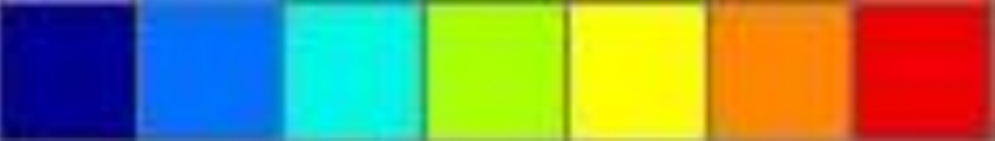}}{B}) in both accuracy and time for \texttt{summary} tasks.

\revised{

\paragraph{\textbf{Takeaways: Design Guidelines.}}
Theoretical work provides concrete rules that allow visualization recommendation algorithms to immediately detect bad encoding choices for a specific data type. In this way, theoretical rules can help visualization recommendation algorithms automatically prune the space of recommendation candidates.
Meanwhile, experimental work examines some hypotheses on how visual encodings perform in practical scenarios. The resulting rankings can inform heuristics-driven visualization recommendation algorithms like Voyager~\cite{Wongsuphasawat2015voyager,Wongsuphasawat2017voyager2} (see~\autoref{subsec:inform-voyager}).
We observe the following takeaways that could inform both human and algorithmic design decisions:
\begin{itemize}[nosep]
    \item In general, position encodings (PX, PY) are the top choices for representing all data types (quantitative, nominal, ordinal).
    \item Ar and CS encodings are also top choices for visualizing quantitative data. 
    \item CH, S, and Ar encodings are effective for visualizing nominal data. However, different combinations have different perceptual discrminabilities. For example, CH is generally more discriminable with filled shapes than with unfilled ones.
    \item CH is not expressive for representing ordinal data, and CS performs better than CH.
\end{itemize}
}

\subsection{Chart Types}
\label{sec:litrev-charts}

Although research findings at the individual encoding level can help systems avoid obvious pitfalls (e.g., choosing a poor color scheme), they fail to clarify \textit{interference effects}~\cite{Kim2018assessing} between encodings.
\revised{For example, the optimal encoding for a single quantitative attribute may not apply when this attribute is also rendered alongside two nominal attributes.}
To address the complexities of combining encodings into full visualization designs, we re-cluster research papers by the chart types they cover.
\revised{We discuss how well existing literature covers different chart types, then summarize observed coverage and study outcomes from the literature towards achieving our two remaining goals: \textit{(1) comparing different variants of a single chart type, and (2) comparing different chart types to identify better visualization designs.}}

\begin{table}
  \caption{Literature coverage for different chart types. The papers are grouped by their category, \textit{theory}, experiment, and \underline{hybrid}.}
  \centering
  \begin{tabular}{L{8.2em}|L{16.4em}}
    \toprule
         & \small \textbf{Relevant Work} \\
        \midrule
        \small \textbf{Scatterplot} & \small   \cite{Heer2010crowdsourcing,Li2010judging,Gleicher2013perception,Albers2014task,Gramazio2014relation,Kanjanabose2015multitask,Correll2017regression,Smart2019measuring,Liu2021data,Lu2021modeling,Hong2021weighted,Harrison2014ranking,Kay2015beyond,Szafir2017modeling,Burlinson2018open,Kim2018assessing,Saket2018task,Mccoleman2021rethinking,Godau2016perception,Mylavarapu2019ranked}, \textit{\cite{Szafir2016four,Sarikaya2017scatterplots}}, \underline{\cite{Micallef2017towards,Wang2018optimizing}} \\
        \cellcolor[HTML]{ededed}\small \textbf{Bar Chart} &  \cellcolor[HTML]{ededed}\small  \cite{Heer2010crowdsourcing,Talbot2014four,Albers2014task,Godau2016perception,Ondov2019face,Srinivasan2018what,Xiong2019biased,Waldner2019comparison,Nothelfer2019measures,Mylavarapu2019ranked,Karduni2020bois,Ceja2020truth,Lu2021modeling,Cleveland1984graphical,Harrison2014ranking,Kay2015beyond,Szafir2017modeling,Saket2018task,Kosara2019impact,Redmond2019visual,Correll2019truncating,Pena2020comparison,Mccoleman2021rethinking,Zhao2019neighborhood}, \textit{\cite{Szafir2016four}},\underline{\cite{Jardine2020perceptual}}\\
        \small \textbf{Line Chart} & \small  \cite{Gogolou2018comparing,Ondov2019face,Xiong2019biased,Correll2019truncating,Javed2010graphical,Aigner2011bertin,Correll2012comparing,Harrison2014ranking,Kay2015beyond,Albers2014task,Szafir2017modeling,Correll2017regression,Saket2018task,Mccoleman2021rethinking},\textit{\cite{Szafir2016four}} \\
        \cellcolor[HTML]{ededed}\small \textbf{Area Chart} & \small  \cellcolor[HTML]{ededed}\cite{Javed2010graphical,Correll2017regression,Gogolou2018comparing,Harrison2014ranking,Kay2015beyond} \\
        \small \textbf{Pie/Donut Chart} & \small \cite{Heer2010crowdsourcing,Skau2016arcs,Waldner2019comparison,Lu2021modeling,Harrison2014ranking,Kay2015beyond,Ondov2019face,Saket2018task,Kosara2019impact,Redmond2019visual} \\
        \cellcolor[HTML]{ededed}\small \textbf{Heatmap} & \cellcolor[HTML]{ededed}\small \cite{Gogolou2018comparing,Correll2012comparing,Albers2014task,Schloss2018mapping} \\
        \small \textbf{Parallel Coordinates} & \small  \cite{Li2010judging,Kanjanabose2015multitask,Harrison2014ranking,Kay2015beyond} \\
        \cellcolor[HTML]{ededed}\small \textbf{Geomap/Cartogram} & \cellcolor[HTML]{ededed}\small  \cite{Pena2019comparison,Pena2020comparison,Gramazio2014relation,Nusrat2018evaluating,Reda2018graphical,Reda2019evaluating,Golebiowska2020rainbow}, \textit{\cite{Szafir2016four}}, \underline{\cite{Gramazio2016colorgorical}} \\
    \bottomrule
  \end{tabular}
  \label{tab:chart-coverage}
\end{table}

\subsubsection{Literature Coverage}

We summarize the chart type coverage by existing literature in~\autoref{tab:chart-coverage}.
Here we use higher-level visualization types to cluster papers.
For example, we lump bubble charts into scatterplots and any variants of area charts like stream graphs into area charts.
We group pie charts and donut charts into one category, as well as geomaps and cartograms.
We can see that bar charts (26/59, $44.07\%$), scatterplots (24/59, $40.68\%$), and line charts (15/59, $25.42\%$) are the top 3 studied chart types, while only a few research papers cover other chart types like area charts (5/59, $8.47\%$), heatmaps (4/59, $6.78\%$), parallel coordinates (4/59, $6.78\%$).
This result is pretty consistent with Beagle~\cite{Battle2018beagle}, based on which bar charts and line charts are the most popular visualization types among all SVG-based visualizations mined from webs. 
On the other hand, we also observe the dearth of theoretical work in the space (see~\autoref{tab:chart-coverage}); thus, we focus on summarizing the study outcomes from empirical work in~\autoref{subsec:within-chart} and~\autoref{subsec:between-chart}. 

\subsubsection{Within Chart Type Comparison}
\label{subsec:within-chart}

\paragraph{Scatterplots.}
Besides the evaluation of the class separability perception~\cite{Smart2019measuring,Wang2018optimizing} (mentioned in~\autoref{subsec:encoding-outcomes}), scatterplots are also examined with different data characteristics~\cite{Kim2018assessing,Gramazio2014relation,Gleicher2013perception}, and marks \cite{Kim2018assessing,Liu2021data,Gleicher2013perception,Hong2021weighted}.
We summarize the findings from these research papers in~\autoref{tab:scatterplot-outcomes}.
Kim \& Heer~\cite{Kim2018assessing} suggest that in general using Ar performs better in \texttt{summary} tasks and CS performs better in \texttt{aggregate} tasks when representing quantitative data; however, the performance exhibits significant variance across different data characteristics (entropy and cardinality).
Gleicher et al.~\cite{Gleicher2013perception} find that higher cardinality (more numbers of points) leads to marginally better performance in \texttt{aggregate} tasks; on the other hand, using redundant encodings (like using the combination of CH and S for a same nominal attribute) would not influence the task performance significantly.
Gramazio et al.~\cite{Gramazio2014relation} suggest that using larger marks can reduce participants' response time; however, Hong et al.~\cite{Hong2021weighted} find that larger and darker marks lead to more bias.
Liu et al.~\cite{Liu2021data} study if the mark orientation would affect task performance and find that the mark orientation that is consistent with the trend of the scatterplots can reduce errors in \texttt{estimate trend} tasks. 

\begin{table}
    \caption{Summarized outcomes for \textbf{scatterplots}. Designs would affect the performance of corresponding tasks and metrics. \revised{One needs to be cautious about using designs (\faWarning) since the effectiveness of visualizations changes dramatically depending on data characteristics.} }
    \centering
    \begin{tabular}{C{12em}C{4em}C{4em}C{2em}}
    \toprule 
    \small \textbf{Designs} & \small \textbf{Tasks} & \small \textbf{Metrics} & \small \textbf{Ref.}\\
    
    \midrule
    \small \raisebox{-.5\height}{\includegraphics[width=0.05\textwidth]{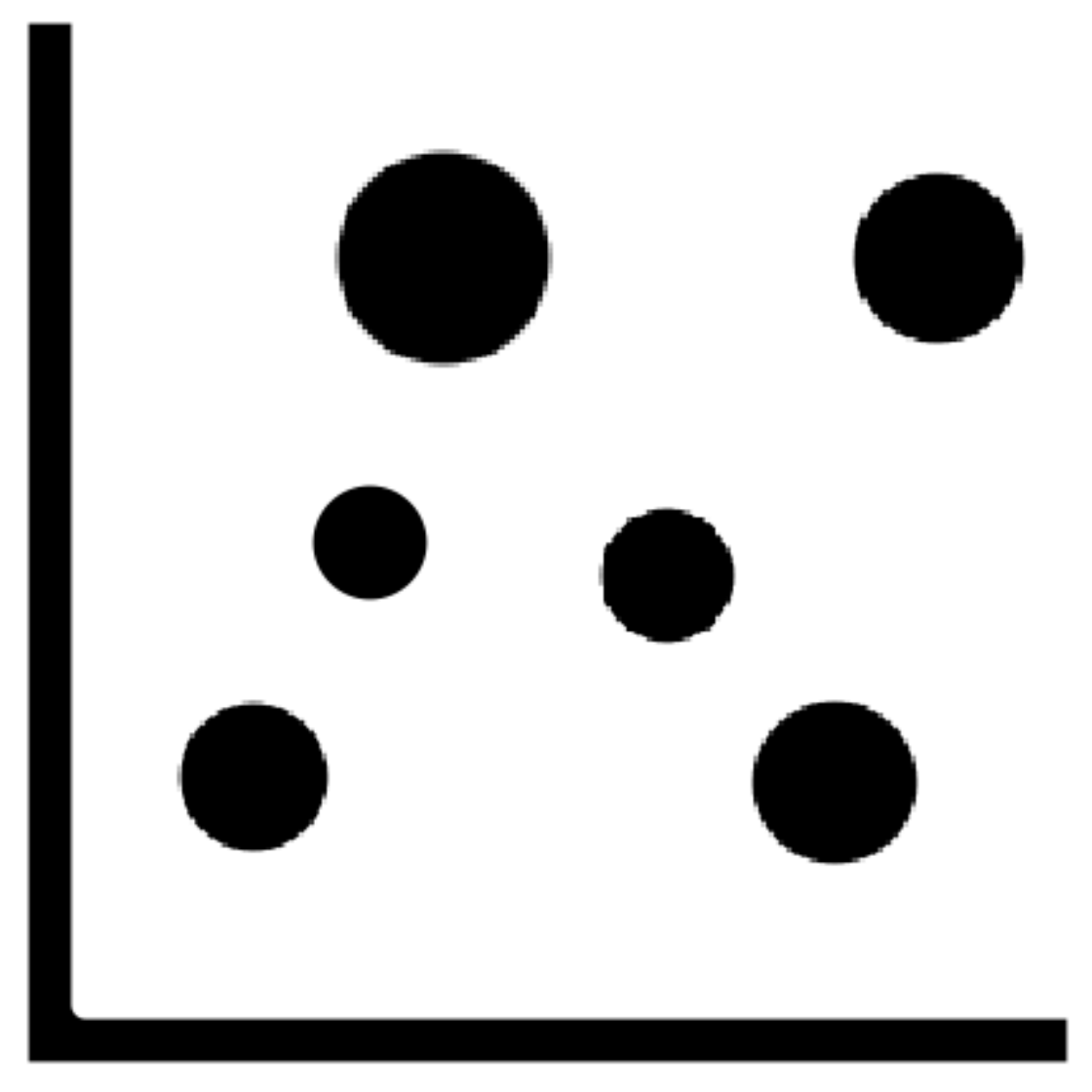}} $>$ \raisebox{-.5\height}{\includegraphics[width=0.05\textwidth]{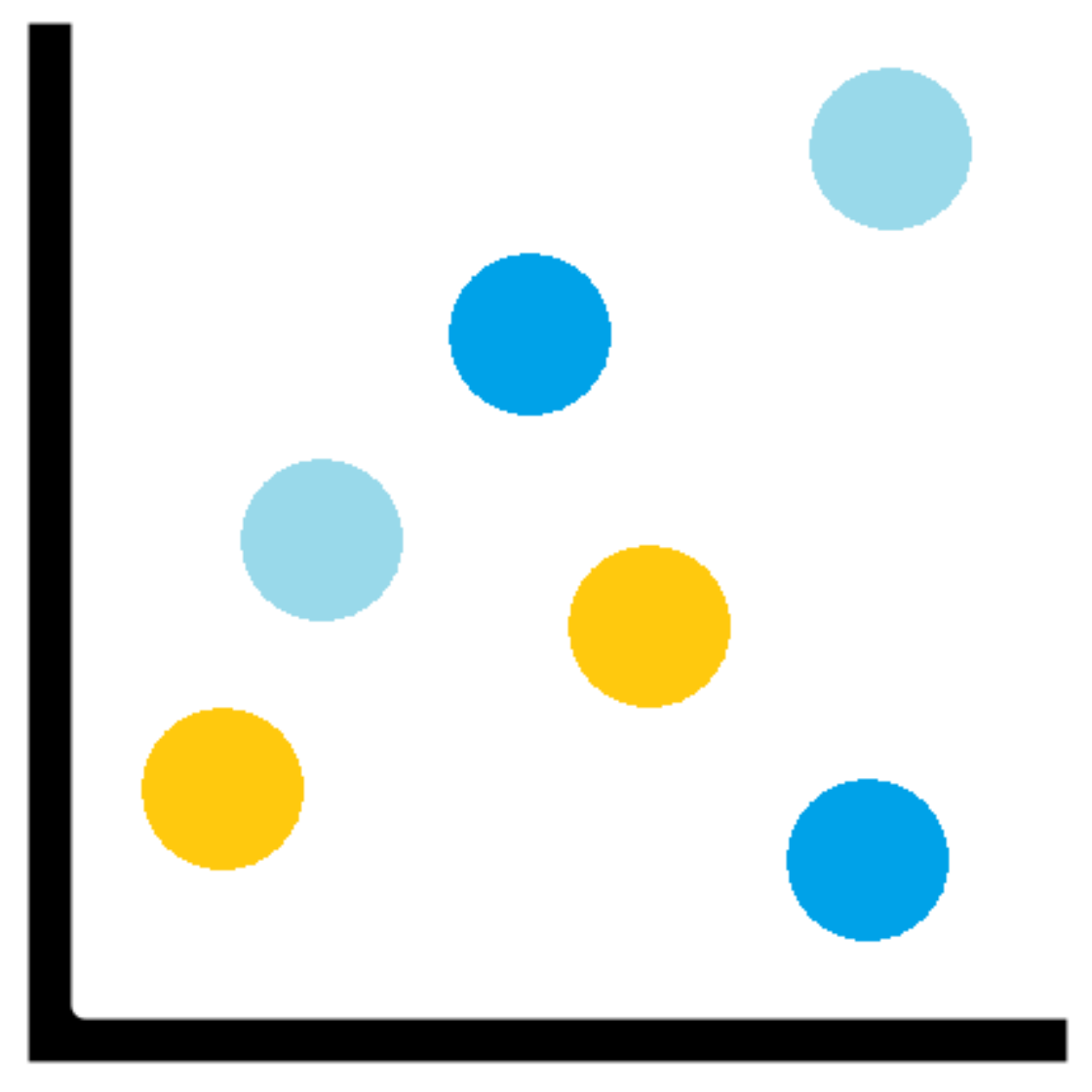}} & \small summary & \small time, accuracy \\
    
    \small \raisebox{-.5\height}{\includegraphics[width=0.05\textwidth]{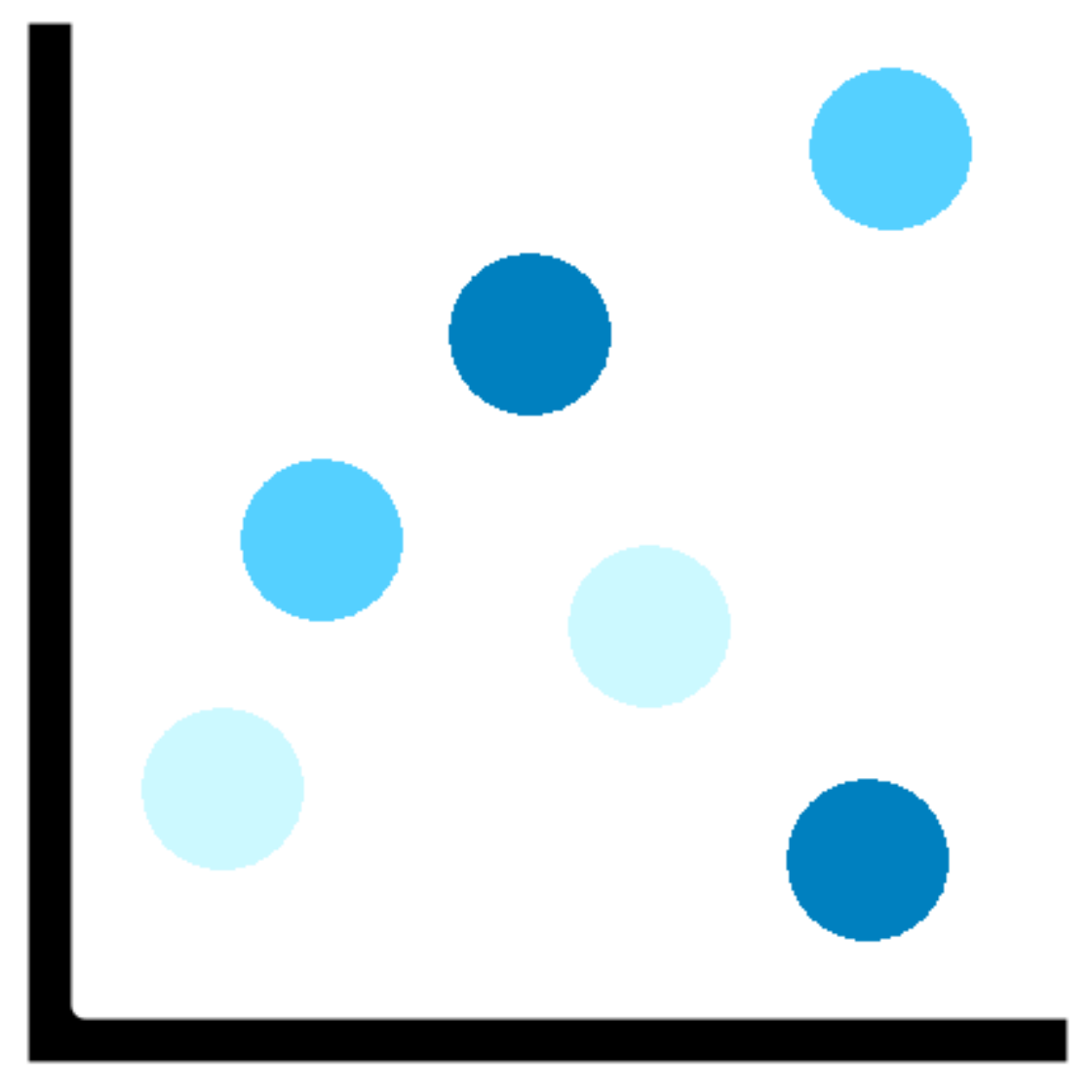}} $>$ \raisebox{-.5\height}{\includegraphics[width=0.05\textwidth]{icons/scatter-ch.pdf}} & \small aggregate & \small accuracy & \small \cite{Kim2018assessing} \\
    
    \small \raisebox{-.5\height}{\includegraphics[width=0.05\textwidth]{icons/scatter-area.pdf}}~ \raisebox{-.5\height}{\includegraphics[width=0.05\textwidth]{icons/scatter-cs.pdf}}~ \raisebox{-.5\height}{\includegraphics[width=0.05\textwidth]{icons/scatter-ch.pdf}} (\faWarning) & \small summary, value & \small time, accuracy &\\
    
    \midrule
    \small \raisebox{-.5\height}{\includegraphics[width=0.05\textwidth]{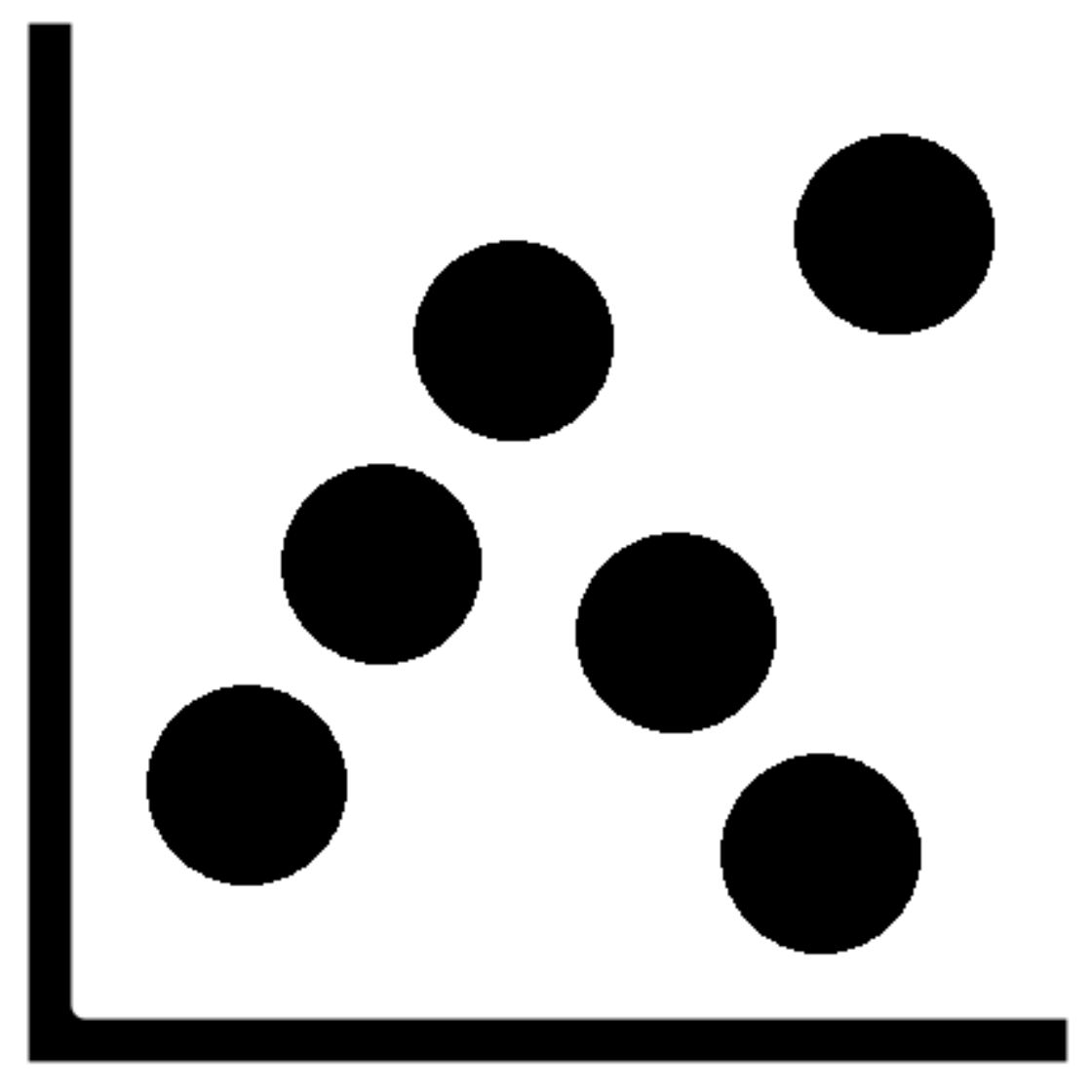}} $>$ \raisebox{-.5\height}{\includegraphics[width=0.05\textwidth]{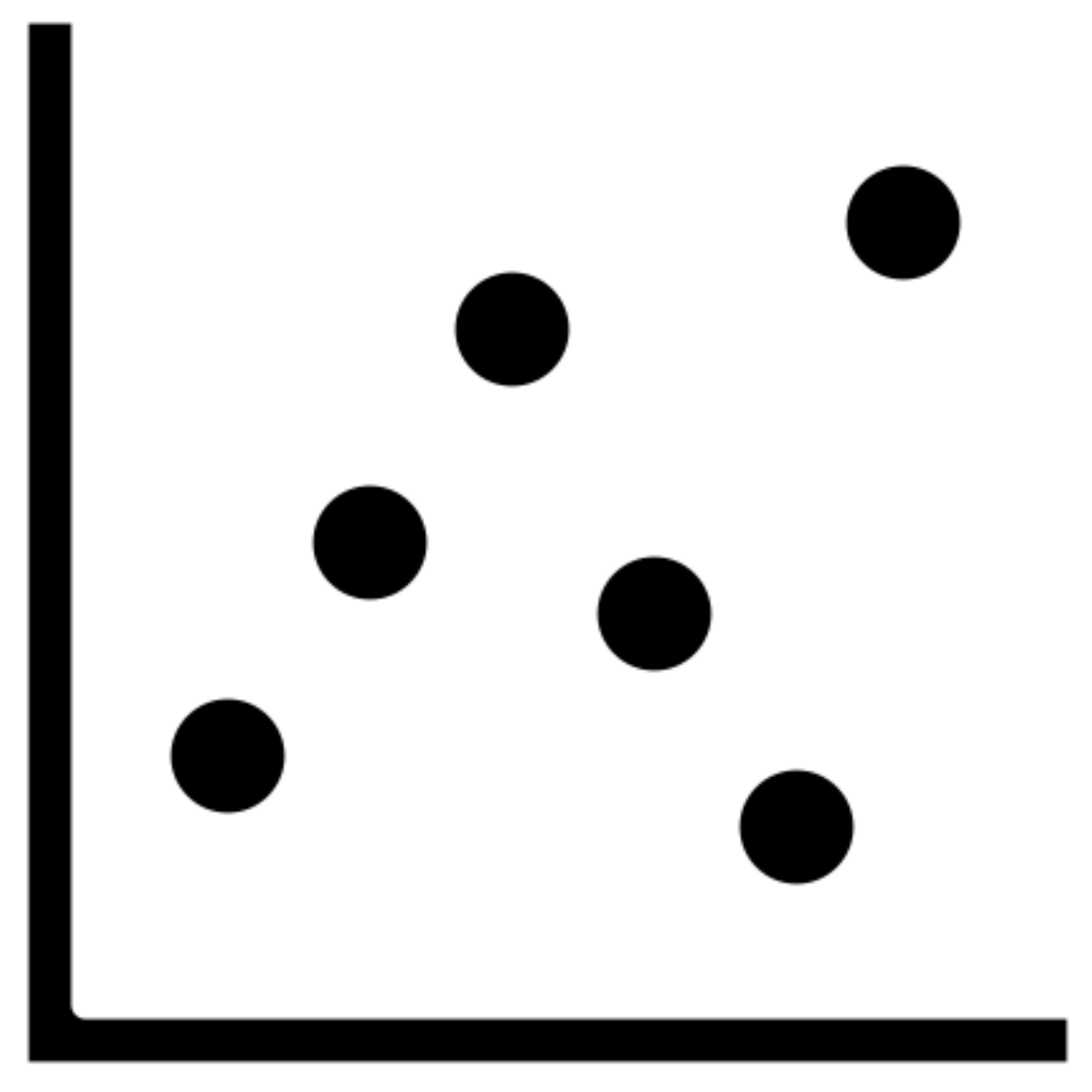}} & \small summary & \small  time & \small \cite{Gramazio2014relation} \\

    \midrule
    \small \raisebox{-.5\height}{\includegraphics[width=0.05\textwidth]{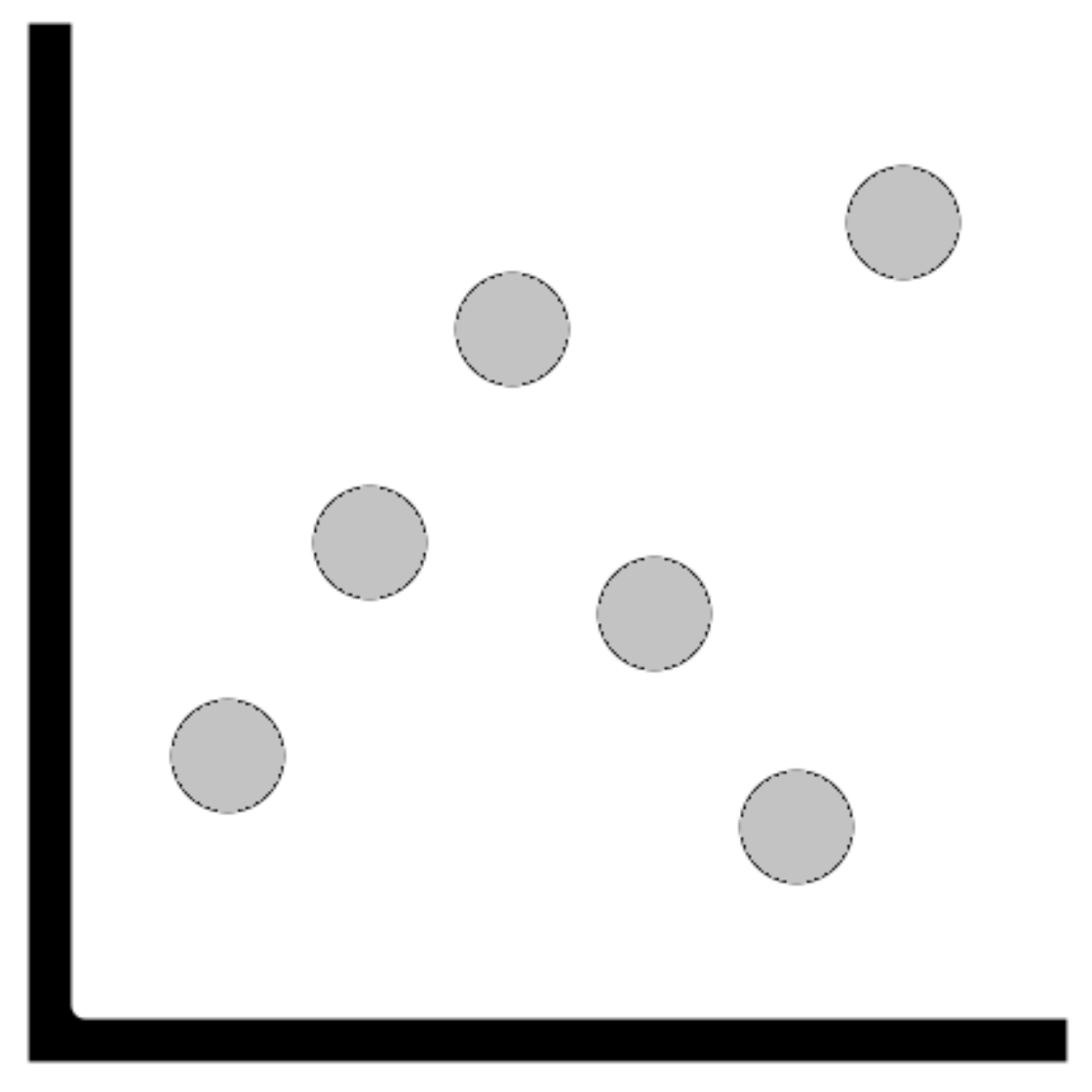}} $>$ \raisebox{-.5\height}{\includegraphics[width=0.05\textwidth]{icons/scatter-larger.pdf}} & \small aggregate & \small bias & \small \cite{Hong2021weighted} \\

    \midrule
    \small \raisebox{-.5\height}{\includegraphics[width=0.05\textwidth]{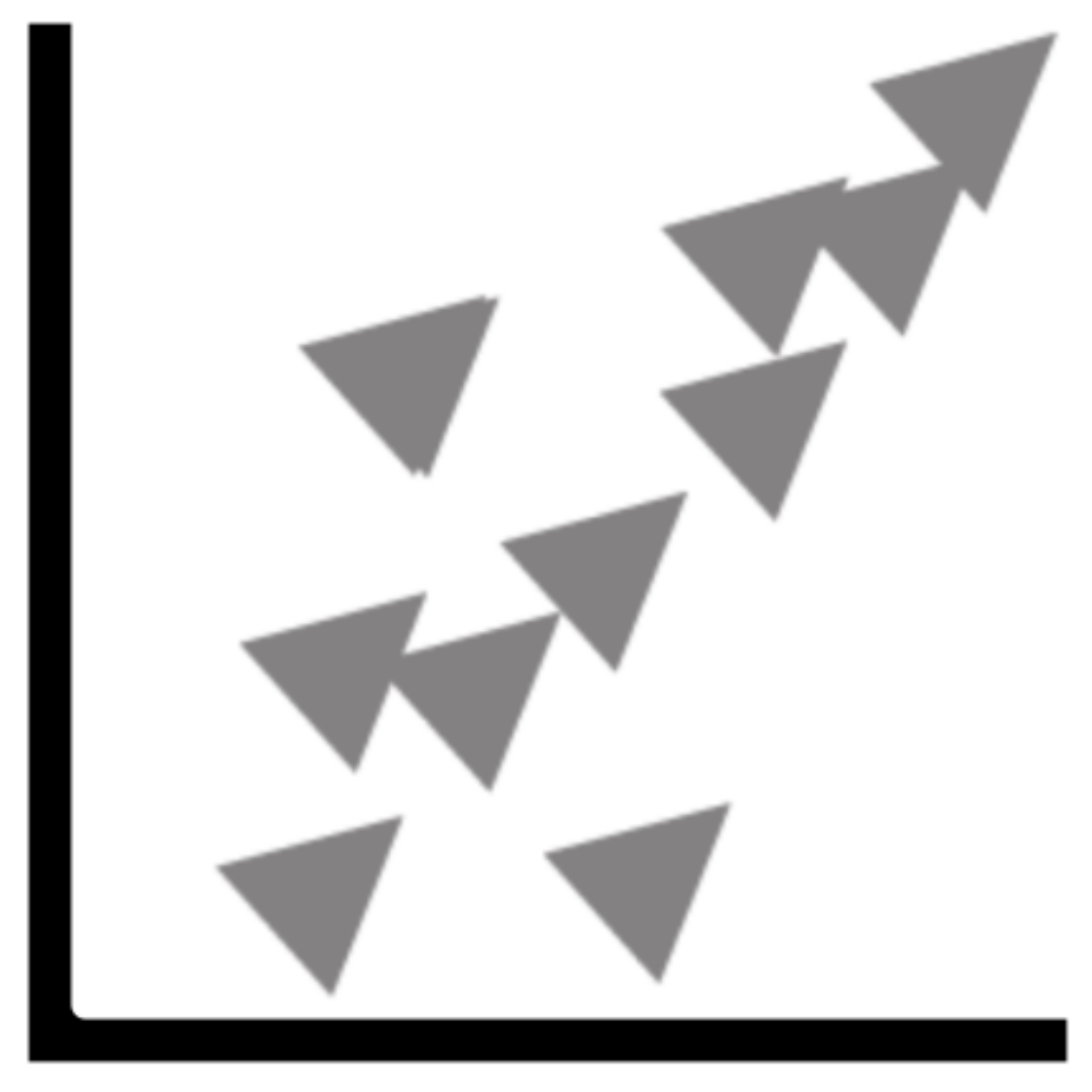}} $>$ \raisebox{-.5\height}{\includegraphics[width=0.05\textwidth]{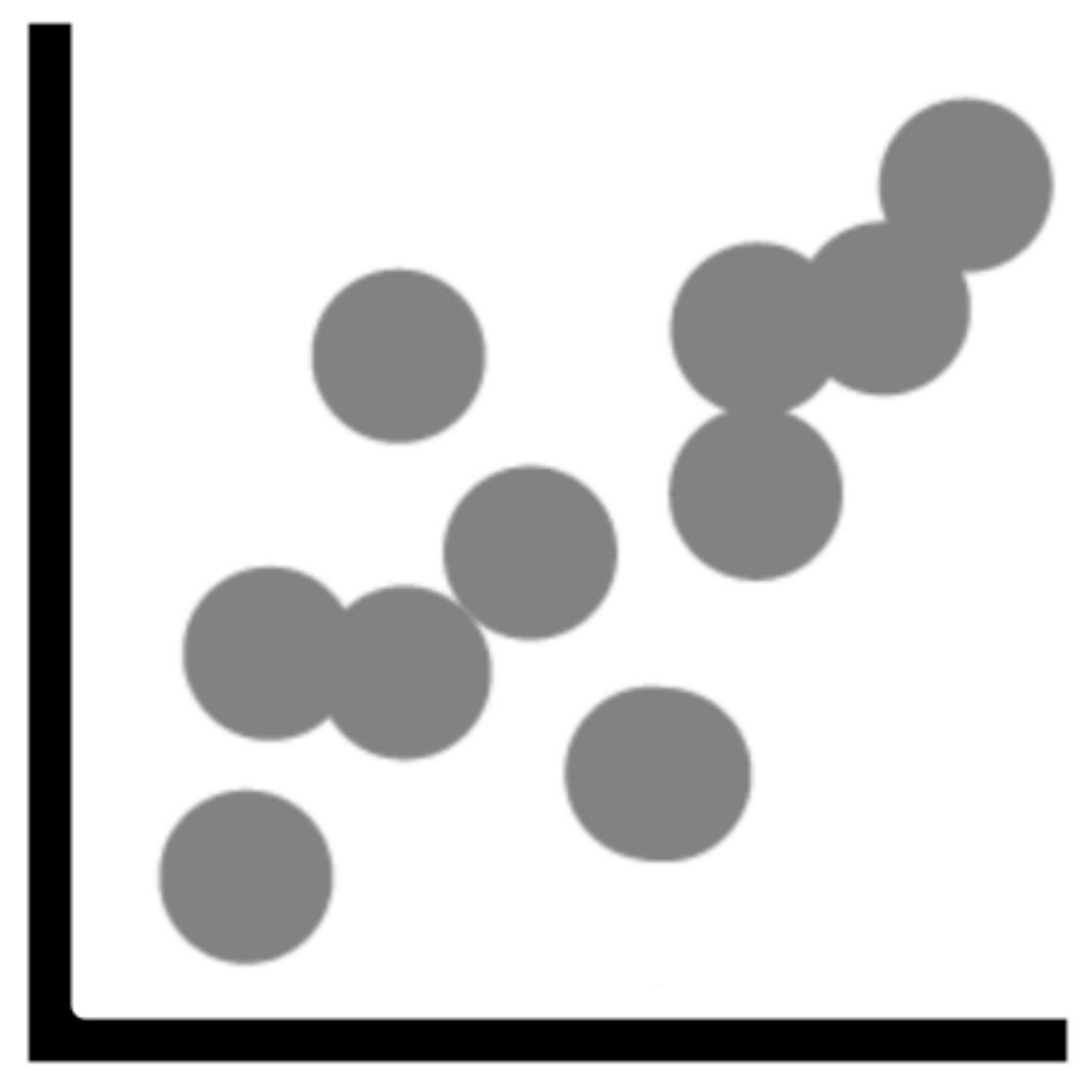}} & \small correlate
    & \small accuracy & \small \cite{Liu2021data} \\

    \midrule
    \small \raisebox{-.5\height}{\includegraphics[width=0.05\textwidth]{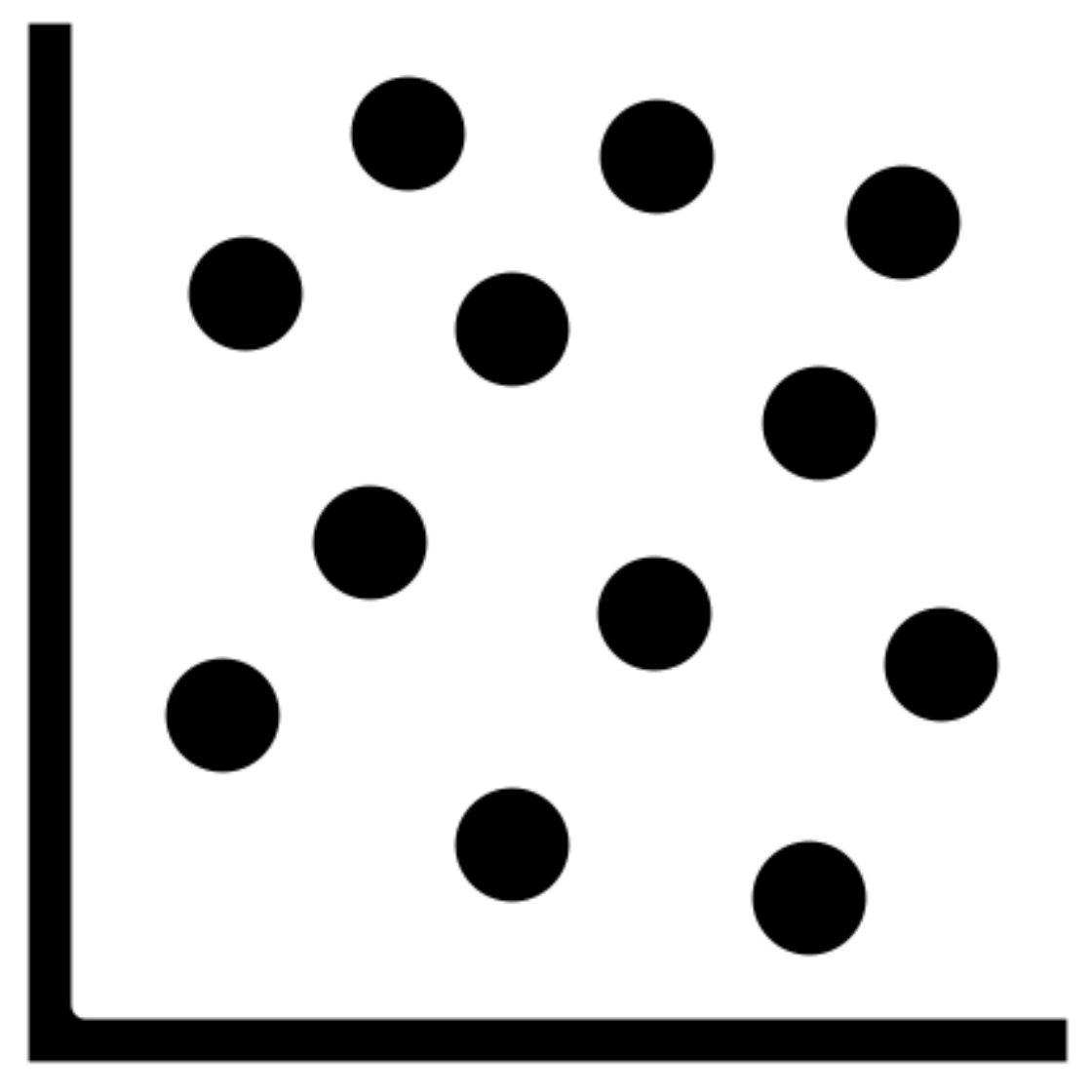}} $>$ \raisebox{-.5\height}{\includegraphics[width=0.05\textwidth]{icons/scatter-less.pdf}} & & & \\

    \small \raisebox{-.5\height}{\includegraphics[width=0.05\textwidth]{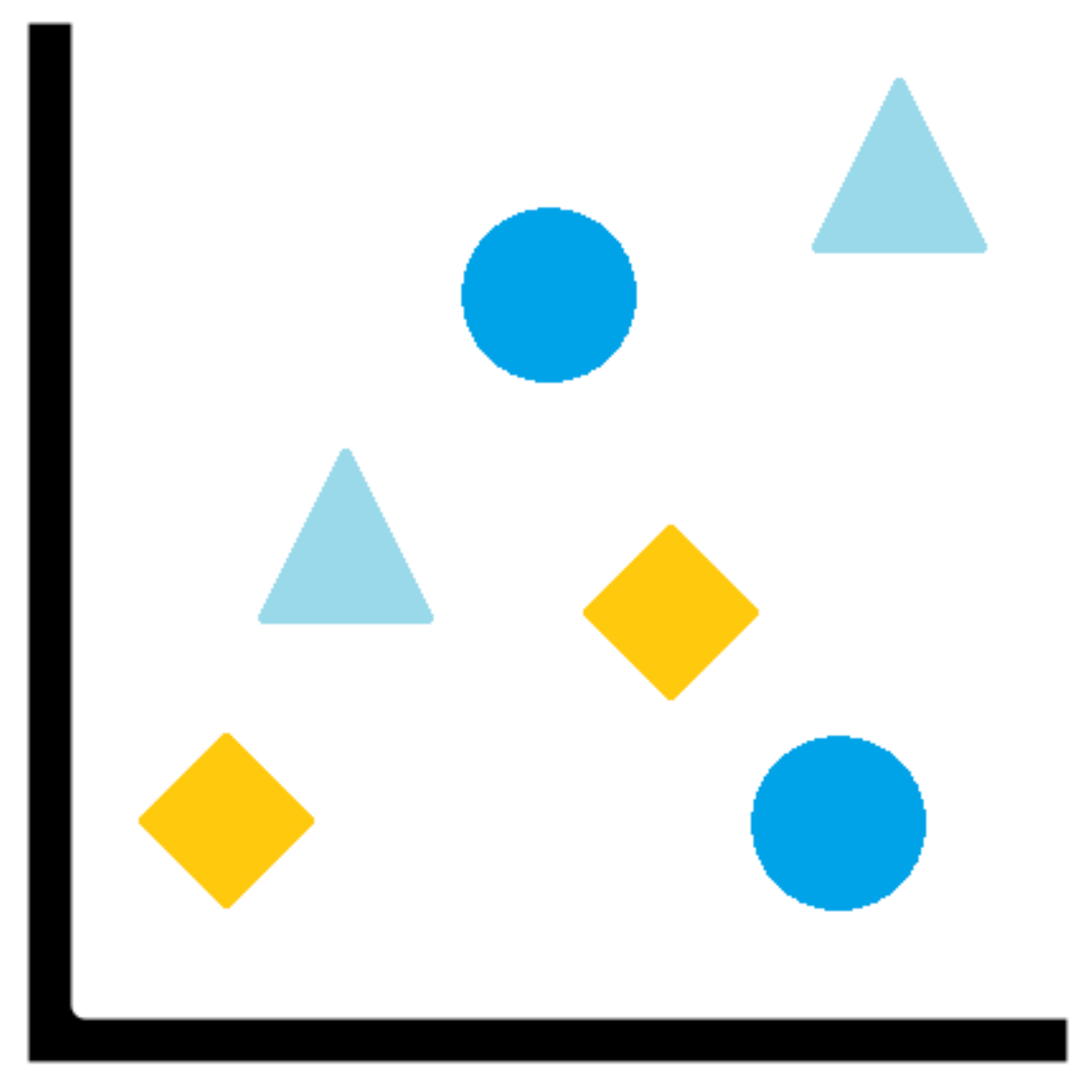}} $=$ \raisebox{-.5\height}{\includegraphics[width=0.05\textwidth]{icons/scatter-ch.pdf}} & \multirow{-2}{*}{\small aggregate} & \multirow{-2}{*}{\small accuracy} & \multirow{-2}{*}{\small\cite{Gleicher2013perception}}\\
    
    \bottomrule
    \end{tabular}
    \label{tab:scatterplot-outcomes}
\end{table}

\paragraph{Bar Charts.}
Unlike scatterplots, bar charts are often studied with different variants~\cite{Ceja2020truth,Karduni2020bois,Srinivasan2018what,Nothelfer2019measures,Talbot2014four} and arrangements~\cite{Talbot2014four,Zhao2019neighborhood}.
Srinivasan et al.~\cite{Srinivasan2018what} and Nothlfer \& Franconeri~\cite{Nothelfer2019measures} evaluate different bar chart variants for \textbf{comparing changing data}.
They both find that visualizing data differences yields better performance and suggest using charts with difference overlays since only visualizing deltas would lose the context of base values.
For \textbf{visualizing disproportionate values}, Karduni et al.~\cite{Karduni2020bois} propose using Du Bios wrapped bar charts.
They find that wrapped bar charts lead to higher accuracy but sometimes at the cost of more time needed than basic bar charts.
Other experiments focus more on the \textbf{perception} of bar charts. 
Talbot et al.~\cite{Talbot2014four} explore variations of bar charts originally studied by Cleveland \& McGill~\cite{Cleveland1984graphical} and find that shorter bars are more difficult for \texttt{sort} tasks. 
Zhao et al.~\cite{Zhao2019neighborhood}, on the other hand, investigate whether neighborhood would influence the perception of bars with \texttt{sort} tasks.
The results show that neighborhood does have an effect, but the effect size is small; other factors like data characteristics have dominated effects. 
Godau et al.~\cite{Godau2016perception} find consistent underestimations with bar charts, which are not affected by the height of bars; moreover, the bias persists even adding a numerical scale or outliers.
Ceja et al.~\cite{Ceja2020truth} recently find that bars with wide aspect ratios are overestimated, bars with tall ratios are underestimated, and bars with square ratios show no systematic bias.

\paragraph{Line Charts.}
We only find one paper that studies line charts solely.
Aigner et al.~\cite{Aigner2011bertin} evaluate three types of line charts (juxtaposition on linear scale, superimposition on log scale, and indexing) with various tasks. 
They find that using indexing generally yields higher accuracy and user preference than the two other types; the advantages with completion time are less clear, although some benefits are shown.

\paragraph{Small Multiples.}
Both Ondov et al.~\cite{Ondov2019face}, and Jardine et al.~\cite{Jardine2020perceptual} study how different arrangements of small multiples would affect the task performance.
In their experiments, five arrangements (stacked, adjacent, mirrored, overlaid, animated) are tested with three chart types---bar charts (with \texttt{find extremum}, \texttt{correlate}, \texttt{determine range}, \texttt{aggregate} tasks), line and donut charts (with \texttt{find extremum} task only).
The results suggest that it is unlikely to discover an easy guideline that specifies the best arrangement or encoding for a given task.

\revised{
\paragraph{\textbf{Takeaways: Research Challenges.}}
Our summary tables provide not only an overview of the current literature studying different chart types but also design guidelines that can be applied by visualization designers and visualization recommendation systems to generate effective visualizations. For example, using CS rather than Ar to represent quantitative data might improve the performance of scatterplots in \texttt{aggregate} tasks, and using redundant encodings might not provide any additional benefit (in~\autoref{tab:scatterplot-outcomes}).
However, we also find some limitations in the existing literature:
\begin{itemize}[nosep]
    \item The literature tends to focus more on studying variants of \textit{scatterplots} and \textit{bar charts} than other chart types. We suggest the community study more variants of the underexplored chart types to better understand the interference effects between different encoding channels.
    \item Although the \textit{line chart} is one of the most popular chart types used on the web~\cite{Battle2018beagle}, not many evaluations of variants exist; thus, we urge more experiments examining different variations of line charts. 
    \item We also find that existing study results differ based on individual factors like data characteristics, tasks, experiment setups, and participants. We suggest more experiments and theories for concluding which visualization design to pick under different analysis scenarios.
\end{itemize}

}

\subsubsection{Between Chart Type Comparison}
\label{subsec:between-chart}

\paragraph{Time Series Data.}
We observe that visualizing time series data is often discussed in existing literature~\cite{Javed2010graphical,Correll2012comparing,Albers2014task,Gogolou2018comparing,Pena2019comparison,Pena2020comparison}.
Correll et al.~\cite{Correll2012comparing} perform an empirical experiment of a \texttt{aggregate} task for time series data; four display conditions were tested: ordered/permuted line chart and ordered/permuted colorfield chart.
The results suggest that colorfield charts outperform in accuracy across all difficulty levels.
Albers et al.~\cite{Albers2014task} extend the work of Correll et al. in a follow-on experiment by testing more visualization designs with more tasks.
The results confirm that different designs support different tasks;  position-based charts outperform in some tasks (\texttt{find extremum}, \texttt{determine range}) while color-based charts perform better in others (\texttt{aggregate}, \texttt{find anomalies}).
Instead of testing encoding performance (\textsl{position} vs. \textsl{color}), Javed et al.~\cite{Javed2010graphical} explore user performance for \texttt{find extremum}, \texttt{determine range}, \texttt{sort} tasks for different line graph techniques (shared-space vs. split-space).
They find no significant difference between these two techniques in terms of accuracy; however, shared-space techniques are faster in \texttt{find extremum} while split-space ones are faster in \texttt{sort} tasks.

\begin{table*}
  \caption{Study outcomes from experiments evaluating \textbf{systematic bias}. \revised{In general, there exist systematic bias in bar charts, while one experiment~\cite{Godau2016perception} found overestimates in bars, and another~\cite{Xiong2019biased} found overestimates in bars. A more recent experiment~\cite{Ceja2020truth} found that the systematic bias in bars is related to the aspect ratio of bars.}}
  \centering
  \begin{tabular}{ccccc}
    \toprule
    \small \textbf{Underestimate} & \small \textbf{Overestimate} & \small \textbf{Perception Pull} & \small \textbf{No Bias} & \small \textbf{Ref.} \\
    \midrule
    \raisebox{-.5\height}{\includegraphics[width=0.05\textwidth]{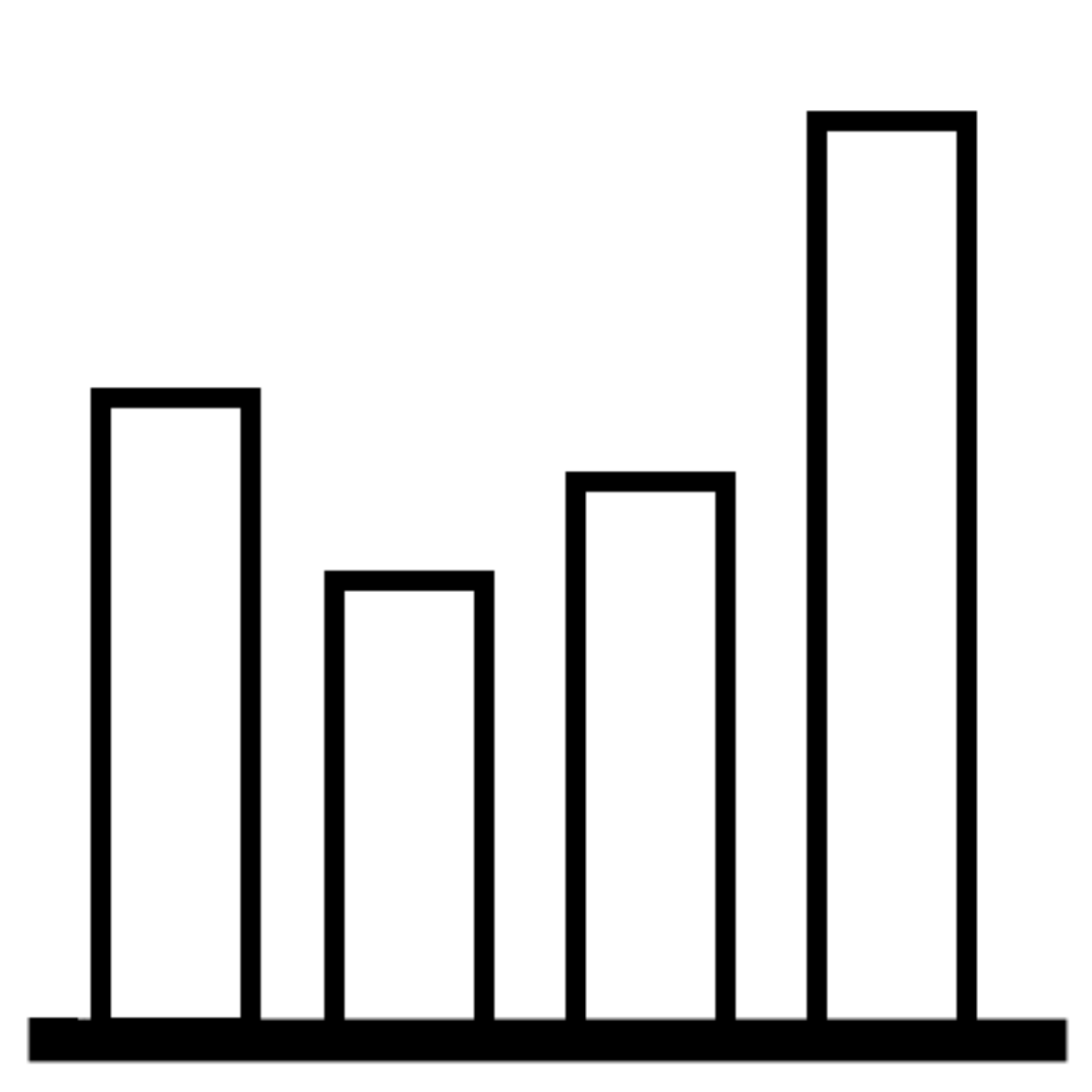}} \small\revised{(bar)}  &  &   & 
     \raisebox{-.5\height}{\includegraphics[width=0.05\textwidth]{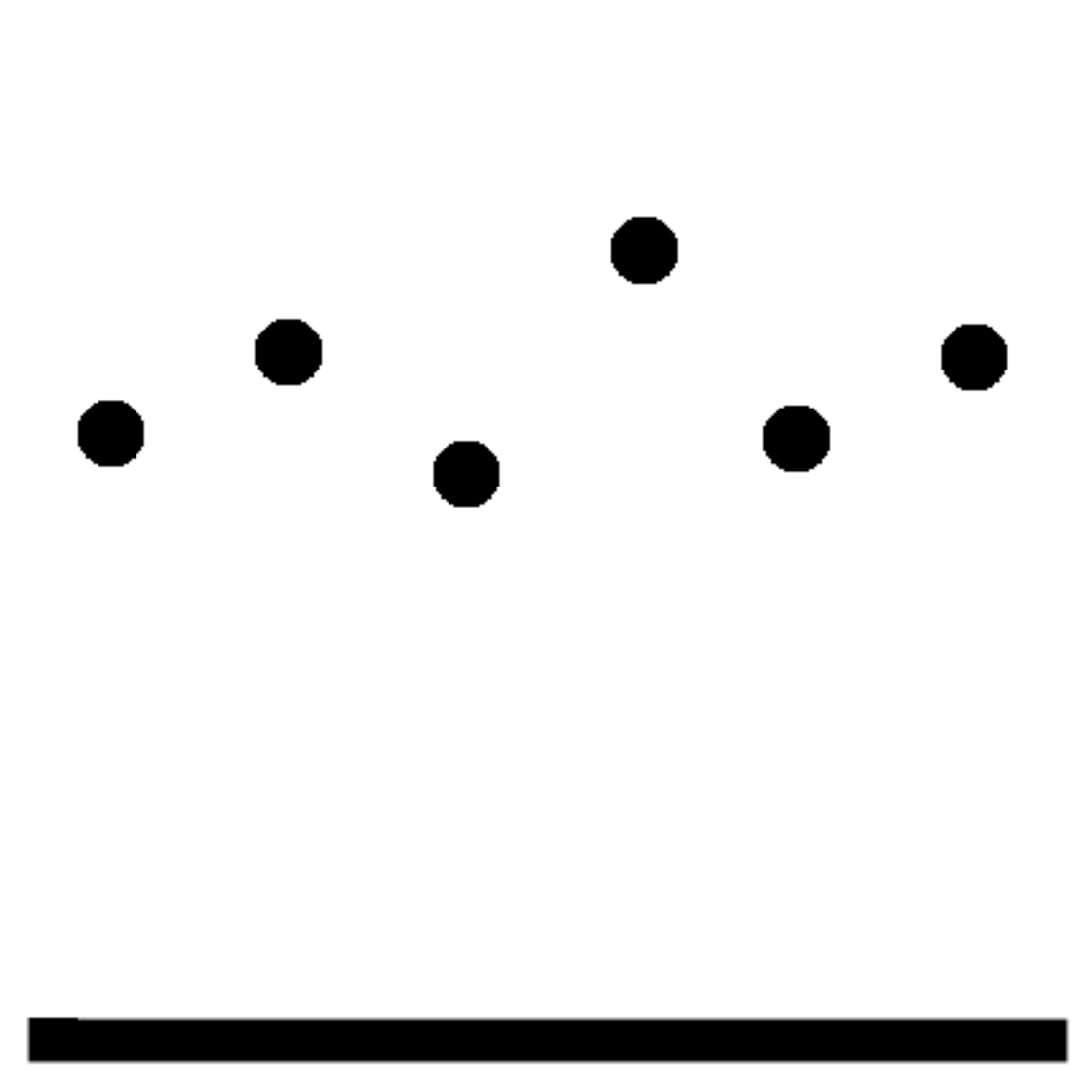}} \small\revised{(point)} & \small \cite{Godau2016perception} \\
     \midrule
      \raisebox{-.5\height}{\includegraphics[width=0.05\textwidth]{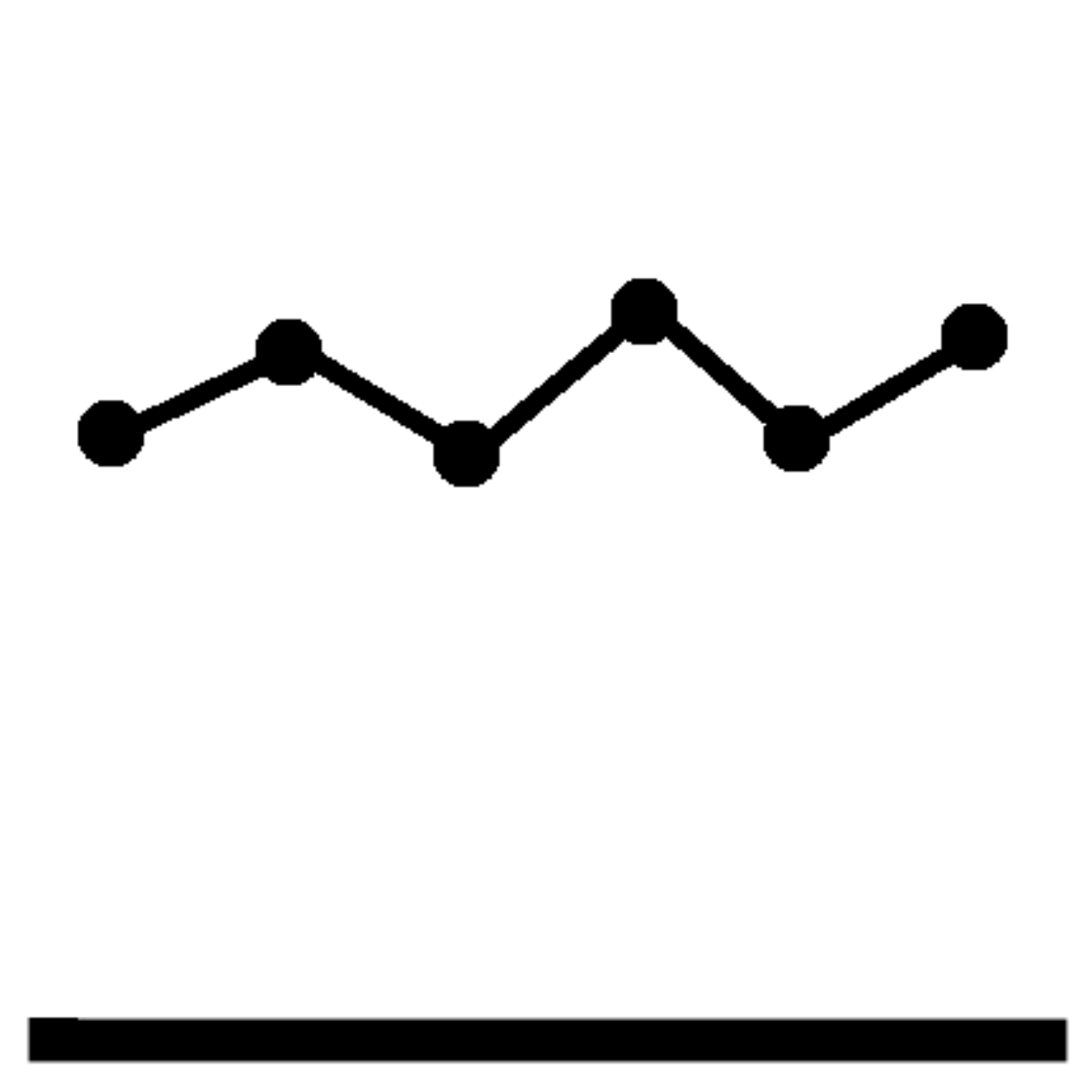}} \small\revised{(line)}  & \raisebox{-.5\height}{\includegraphics[width=0.05\textwidth]{icons/bar.pdf}} \small\revised{(bar)} & 
      \raisebox{-.5\height}{\includegraphics[width=0.05\textwidth]{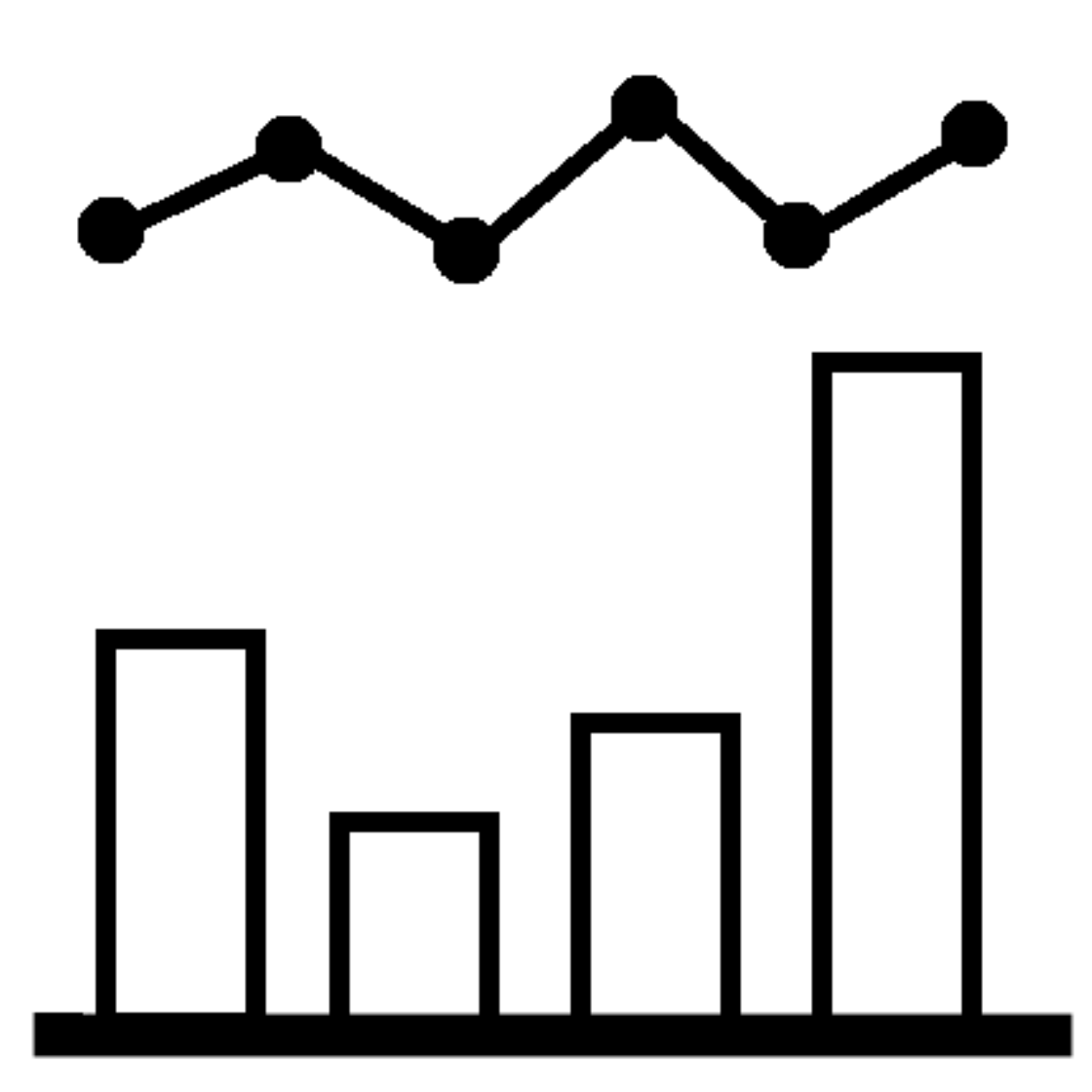}} \small\revised{(line \& bar)} &   & \small \cite{Xiong2019biased} \\
      \midrule
      \raisebox{-.5\height}{\includegraphics[width=0.05\textwidth]{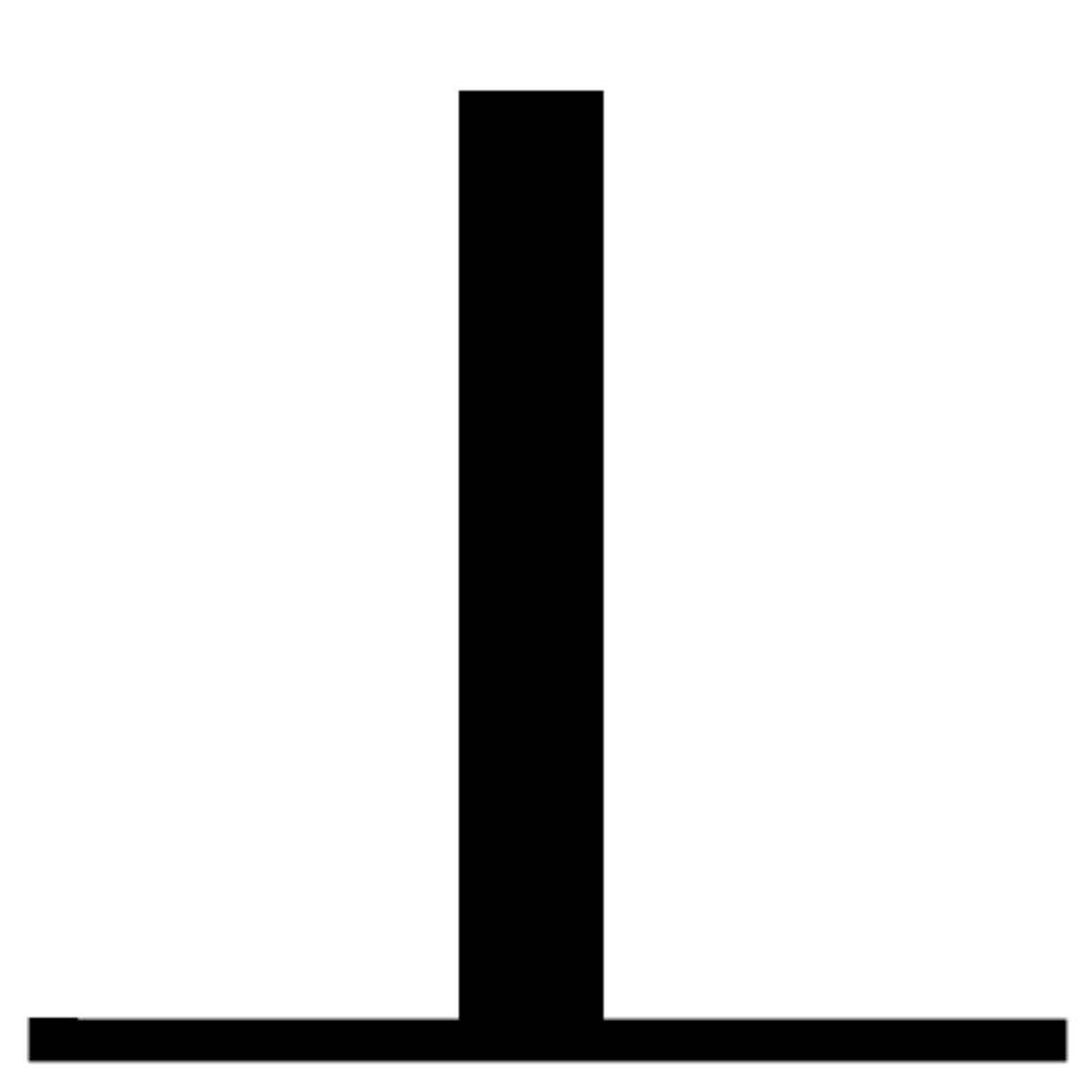}} \small\revised{(tall bars)} &
      \raisebox{-.5\height}{\includegraphics[width=0.05\textwidth]{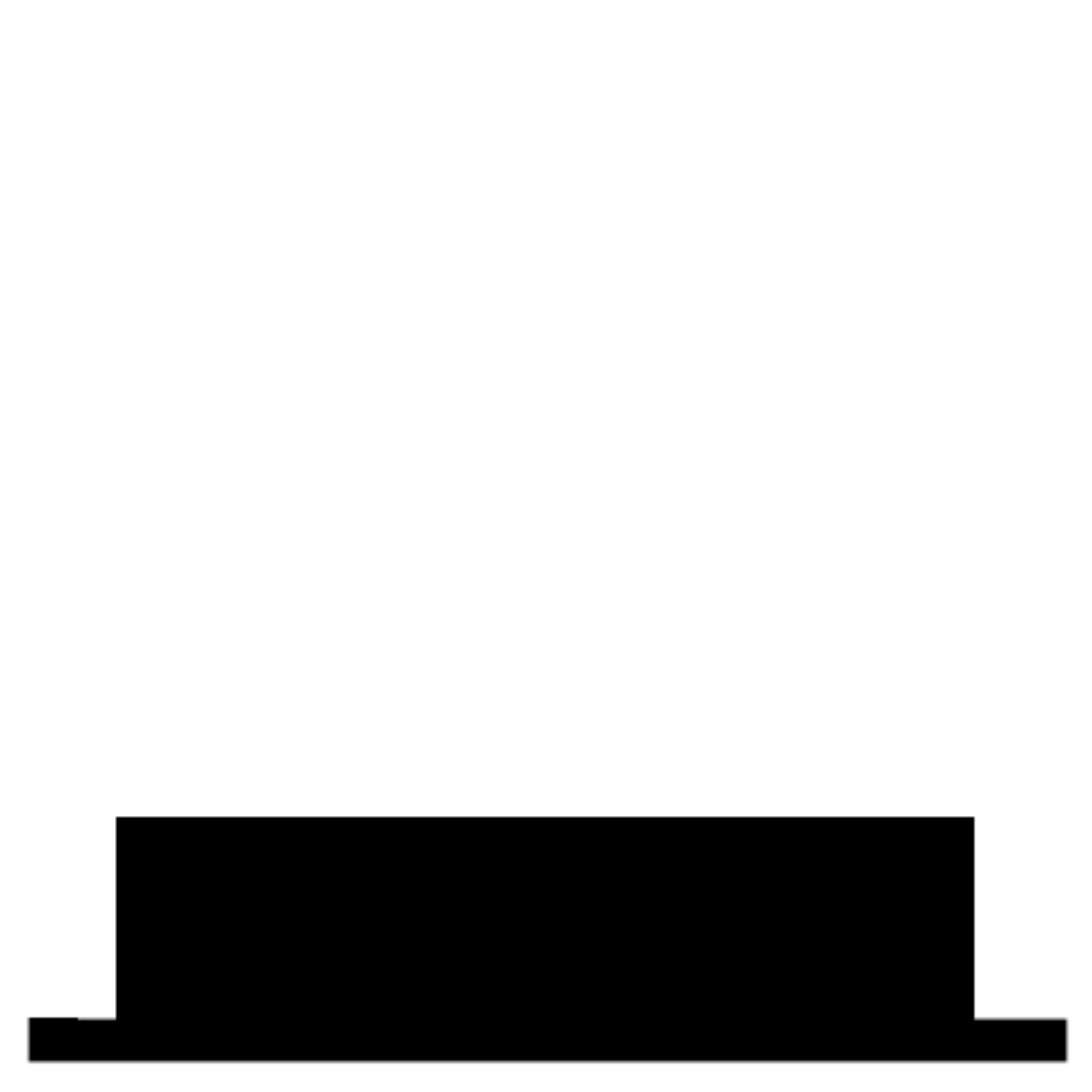}} \small\revised{(wide bars)} &
       &
      \raisebox{-.5\height}{\includegraphics[width=0.05\textwidth]{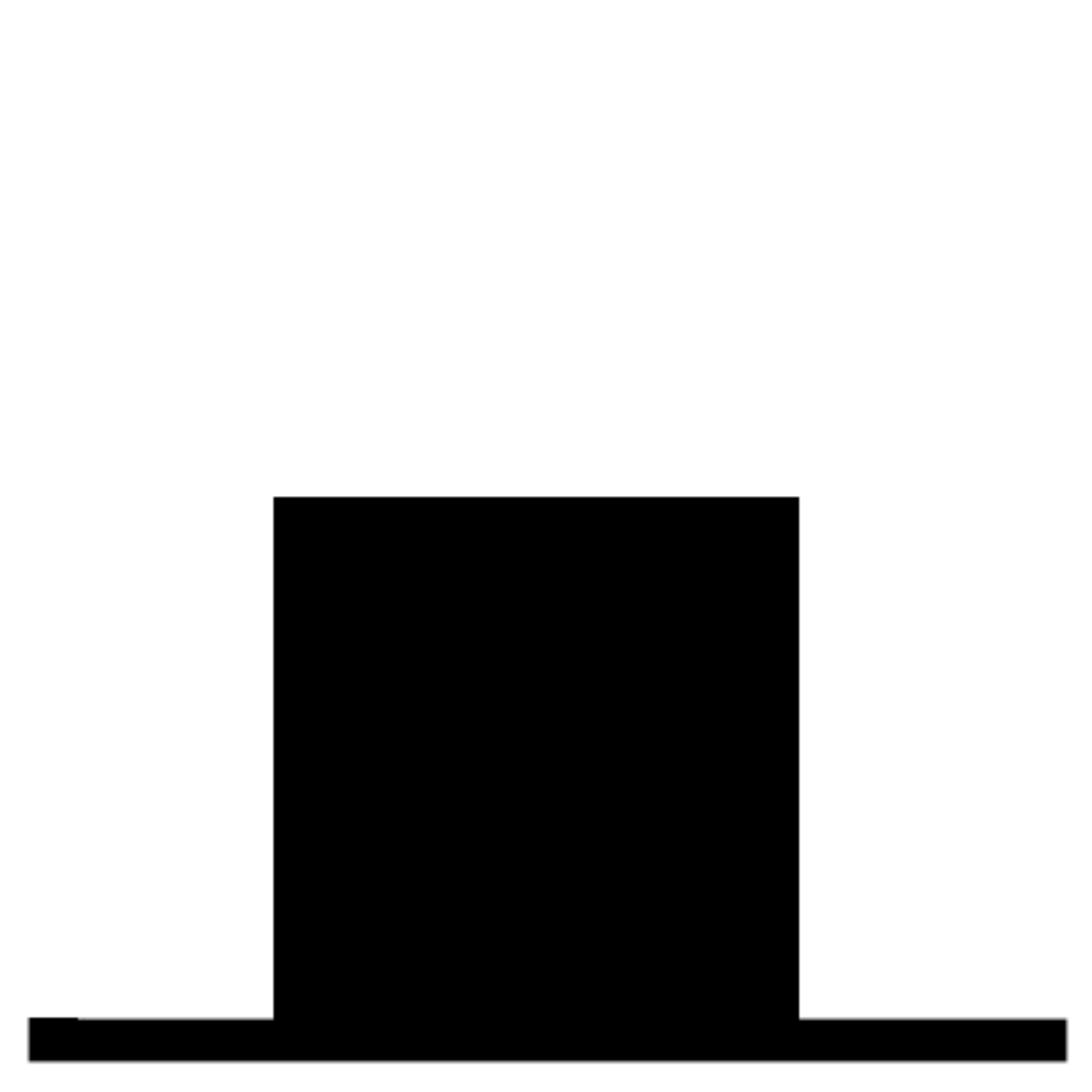}} \small\revised{(square bars)} & \small \cite{Ceja2020truth} \\
    \bottomrule
  \end{tabular}
  \label{tab:bias}
\end{table*}

\paragraph{Systematic Bias.}

Bias evaluation has attracted more attention recently~\cite{Xiong2019biased,Godau2016perception,Ceja2020truth,Correll2019truncating}.
Godau et al.~\cite{Godau2016perception} test whether there is a bias in the central tendency perceived in bar charts, and they find that the mean value is systematically underestimated in bar charts (but not in scatterplots). 
Their other experiments also confirm that the underestimation of the average persists with varying bar heights or adding outliers.
However, Xiong et al.~\cite{Xiong2019biased} reach a completely different conclusion.
They conducted three empirical studies to investigate position perception bias with visualizations containing a single bar/line, multiple bars/lines, and one line with one set of bars. 
In contrast to the results of Godau et al., they found that the perceived average position was significantly biased in both single line charts and single bar charts. 
Line positions are consistently underestimated, while bar positions are overestimated. In the experiments involving multiple data series (multiple lines and/or bars), they also observe an effect of ``perceptual pull'', where the average position estimate for each series was ``pulled'' toward the other.
Aiming to explain this contradiction, recent research by Ceja et al.~\cite{Ceja2020truth} finds that the systematic bias in bars is related to the aspect ratio of bars. 
No systematic bias is shown with square bars, while wide bars are overestimated, and tall bars are underestimated.
We summarize the study outcomes in~\autoref{tab:bias}.

\paragraph{Scatterplots vs. Parallel Coordinates.}

Several current works focus on comparing scatterplots with parallel coordinates~\cite{Li2010judging,Kanjanabose2015multitask,Harrison2014ranking,Kay2015beyond}. 
Li et al.~\cite{Li2010judging} focus on studying \texttt{correlate} task performance between scatterplots and parallel coordinates and find that the degree of correlation between attributes is underestimated in parallel coordinates, suggesting that scatterplots are better options for \texttt{correlate} tasks.
Kanjanabose et al.~\cite{Kanjanabose2015multitask} also perform experiments comparing scatterplots and parallel coordinates but focus on other visual analysis tasks, including \texttt{retrieve value}, \texttt{cluster}, \texttt{find anomalies} and \texttt{determine range}.
The results suggest that parallel coordinates outperform in \texttt{accuracy} across \texttt{cluster}, \texttt{find anomalies} and \texttt{determine range} tasks, and in \texttt{completion time} with \texttt{retrieve value} and \texttt{determine range} tasks.

\paragraph{Bar Charts vs. Pie Charts.}

We find five papers~\cite{Cleveland1984graphical,Heer2010crowdsourcing,Waldner2019comparison,Kosara2019impact,Redmond2019visual} comparing bar charts and pie charts.
Cleveland \& McGill~\cite{Cleveland1984graphical} propose an order of encoding channels based on graphical perception but also test parts of this theory through experiments. 
They use bar charts to assess \textsl{position} and \textsl{length} encodings and pie charts for \textsl{angle} encoding.
Heer \& Bostock~\cite{Heer2010crowdsourcing} replicate these experiments but also adjust the experiments to make results between length and angle encodings comparable.
They conduct the experiments with \texttt{sort} tasks, and both of the results suggest that bar charts perform better than pie charts in terms of the accuracy metric.
Later on, Waldner et al.~\cite{Waldner2019comparison} also report that radial charts perform less accurately, efficiently, and preferably than bar charts in many analytical tasks.
Kosara~\cite{Kosara2019impact} also has similar findings (bar $>$ pie) with \texttt{find extremum} tasks.
However, by comparing the performance of pie charts and bar charts with multiple variants, Redmond~\cite{Redmond2019visual} finds that pie charts perform more accurately with \texttt{retrieve value} tasks.

\paragraph{Multi-Chart Comparisons.}

Some experiments involve a large range of chart types~\cite{Saket2018task,Harrison2014ranking,Saket2018beyond,Mylavarapu2019ranked,Correll2017regression,Skau2016arcs}.
Saket et al.~\cite{Saket2018task} conduct an experiment to evaluate the effectiveness of five 2-encoding visualization designs across all ten analysis tasks (mentioned in~\autoref{tab:tasks}): line chart, bar chart, scatterplot, pie chart, and table.
They confirm that no specific visualization outperforms in every task, and suggest using bar charts for \texttt{clustering}, line charts for \texttt{correlation}, and scatterplots for \texttt{finding anomalies}.
Harrison et al.~\cite{Harrison2014ranking} conduct a crowdsourced experiment to evaluate the human perception of \texttt{correlation} among nine commonly used visualization types, like scatterplots, area charts, line charts, bar charts, pie charts, parallel coordinates, etc. 
The results reveal significant differences in the correlation perception across visualizations, and the results also vary significantly across different data characteristics (different correlation coefficients $r$).
Skau \& Kosara~\cite{Skau2016arcs}, on the other hand, compare the effectiveness of pie charts, donut charts, arc charts, and area charts with \texttt{retrieve value} tasks.
The results show no significant difference between pie charts and donut charts in accuracy, and both perform better than arc and area charts.

\begin{table}
    \caption{Visualization types recommended by \textbf{empirical} work for each visual analysis task. Multiple visualizations recommended for the same task might not be comparable.}
    \centering
    \begin{tabular}{ll}
    
    \toprule
    \small\textbf{Tasks} & \small\textbf{Designs} \\

    \midrule
     \small Retrieve Value & \small \raisebox{-.5\height}{\includegraphics[width=0.05\textwidth]{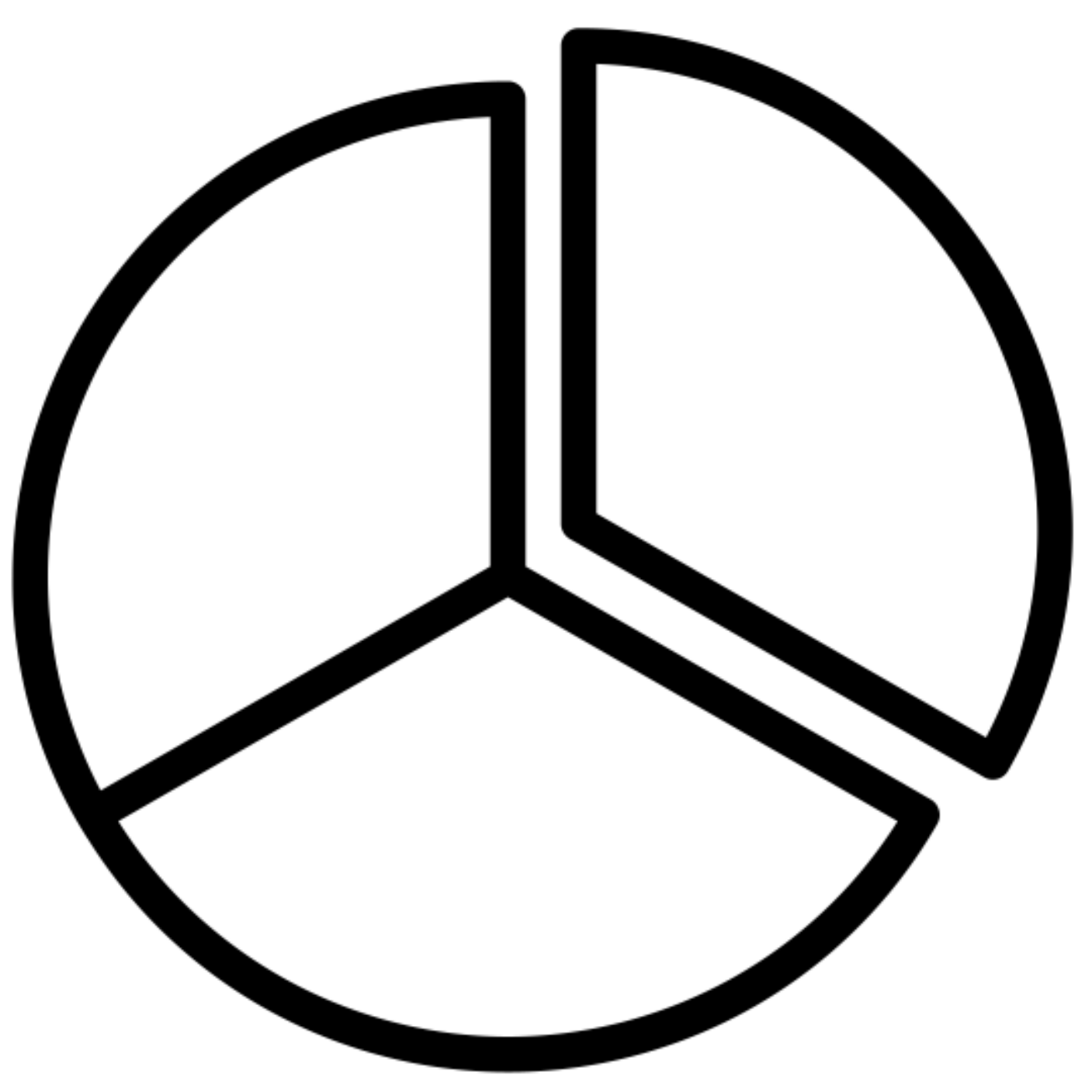}}~\cite{Redmond2019visual},
     \raisebox{-.5\height}{\includegraphics[width=0.05\textwidth]{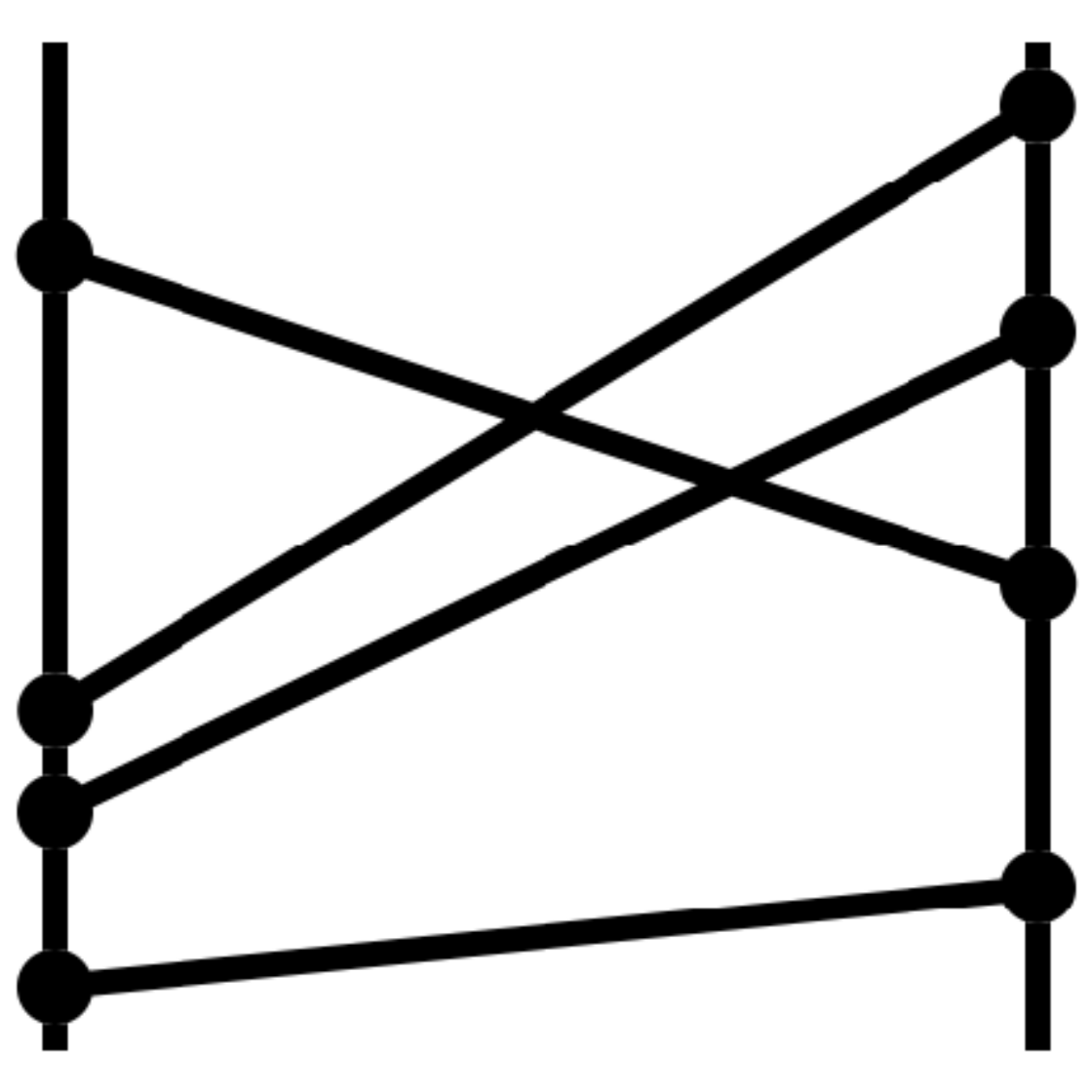}}~\cite{Kanjanabose2015multitask} \\

     \midrule
     \small Filter  & \small \raisebox{-.5\height}{\includegraphics[width=0.05\textwidth]{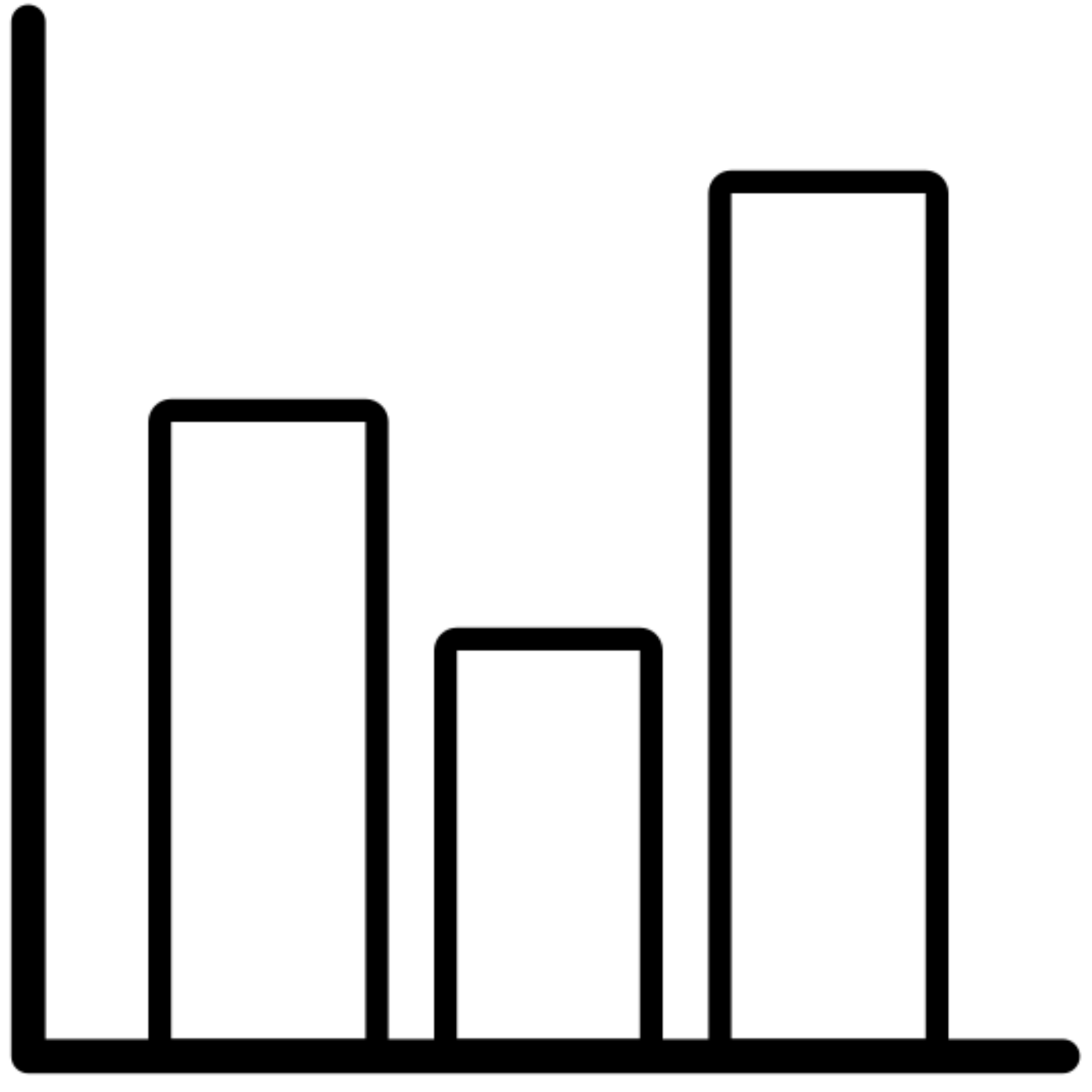}}~\cite{Saket2018task} \\

     \midrule
     \small Sort  & \small \raisebox{-.5\height}{\includegraphics[width=0.05\textwidth]{icons/barchart.pdf}}~\cite{Cleveland1984graphical,Heer2010crowdsourcing,Saket2018task} \\

     \midrule
    \small Cluster & \small  \raisebox{-.5\height}{\includegraphics[width=0.05\textwidth]{icons/barchart.pdf}}~\cite{Saket2018task}, \raisebox{-.5\height}{\includegraphics[width=0.05\textwidth]{icons/pcp.pdf}}~\cite{Kanjanabose2015multitask} \\

    \midrule
     \small Correlate  & \small \raisebox{-.5\height}{\includegraphics[width=0.05\textwidth]{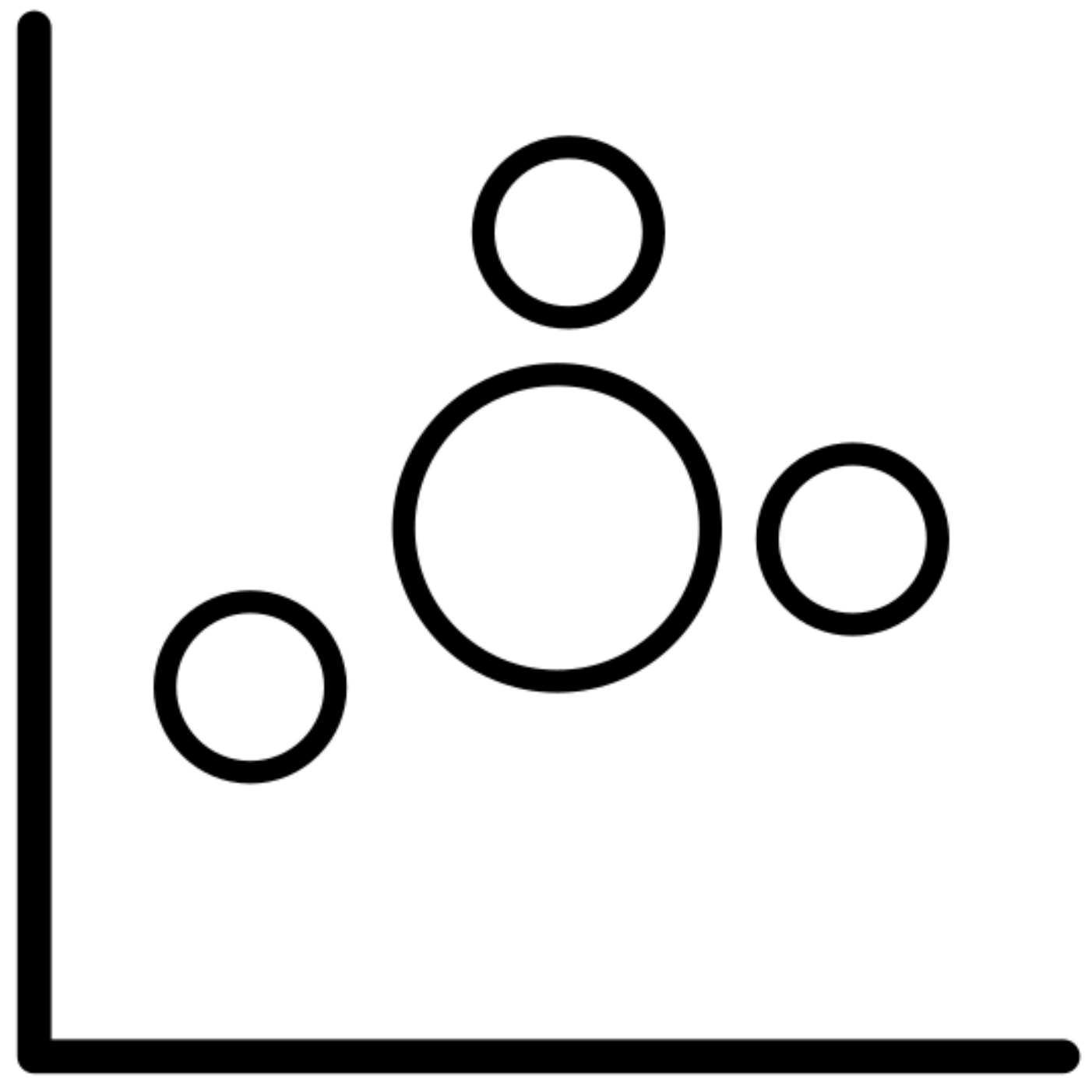}}~\cite{Li2010judging}, \raisebox{-.5\height}{\includegraphics[width=0.05\textwidth]{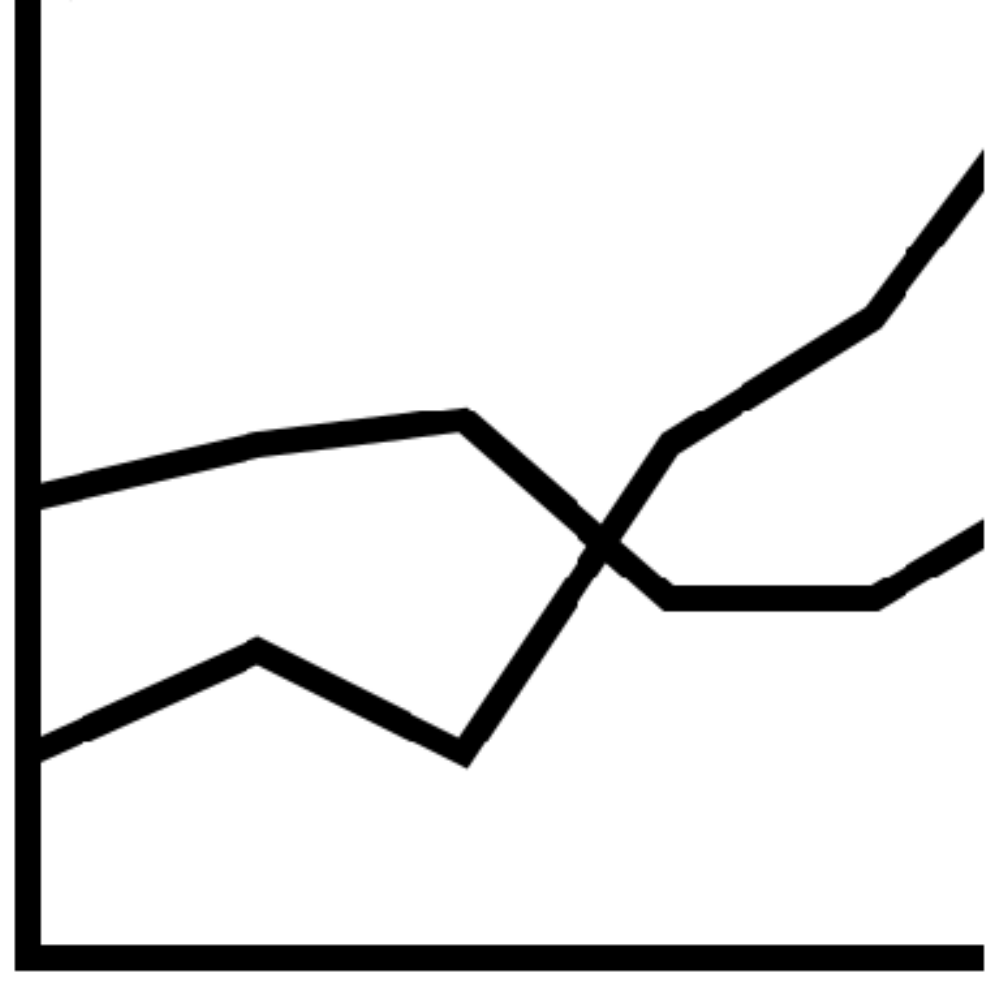}}~\cite{Saket2018task,Correll2017regression} \\

    \midrule
     \small Aggregate & \small \raisebox{-.5\height}{\includegraphics[width=0.05\textwidth]{icons/barchart.pdf}}~\cite{Saket2018task}\\

     \midrule
     \small Find Extremum & \small \raisebox{-.5\height}{\includegraphics[width=0.05\textwidth]{icons/barchart.pdf}}~\cite{Kosara2019impact,Saket2018task} \\

     \midrule
     \small Determine Range & \small \raisebox{-.5\height}{\includegraphics[width=0.05\textwidth]{icons/pcp.pdf}}~\cite{Kanjanabose2015multitask} \\
     
    \midrule
    \small Characterize Distribution  & \small \raisebox{-.5\height}{\includegraphics[width=0.05\textwidth]{icons/barchart.pdf}}~\cite{Saket2018task} \\
    
    \midrule
    \small Find Anomalies & \small \raisebox{-.5\height}{\includegraphics[width=0.05\textwidth]{icons/scatterplot.pdf}}~\cite{Saket2018task},
     \raisebox{-.5\height}{\includegraphics[width=0.05\textwidth]{icons/pcp.pdf}}~\cite{Kanjanabose2015multitask} \\

    \bottomrule
  \end{tabular}
  \label{tab:between-charts-outcomes}
\end{table}

\revised{

\paragraph{\textbf{Takeaways: Design Guidelines.}}
Although we observe gaps in the literature for specific analysis tasks and performance metrics, current studies do provide some guidance on which visualization types work best for common analysis tasks. In \autoref{tab:between-charts-outcomes}, we synthesize these results into concrete design guidelines for ten different tasks:
\begin{itemize}[nosep]
    \item Existing literature suggests using bar charts for the majority of the tasks (six out of ten tasks: \texttt{cluster}, \texttt{filter}, \texttt{sort}, \texttt{distribution}, \texttt{find extremum}, and \texttt{aggregation}). However, some literature~\cite{Xiong2019biased,Godau2016perception,Ceja2020truth} also found that there exist systematic bias in bars. 
    \item Parallel Coordinates are also top choices for four tasks: \texttt{cluster}, \texttt{retrieve value}, \texttt{find anomalies}, \texttt{determine range}.
    \item Currently, no systematic bias is found with point marks~\cite{Godau2016perception}. Scatterplots are also top choices for \texttt{correlate} and \texttt{find anomalies} tasks.
\end{itemize}
}

\section{Discussion: Applications \& Research Challenges}

In this paper, we present a literature review to investigate how visualization designs are compared and ranked in existing theory and experimental work, but also contribute a dataset and synthesize guidelines to facilitate the generation of encoding rules for visualization recommendation systems.
In this section, we show how our contributions could be used directly within these systems.
We also outline the challenges we observed in the literature and suggest research directions in developing new theories and experiments to further encapsulate, enrich and evaluate our understanding of \revised{graphical} perception.

\revised{\subsection{Examples of Using Graphical Perception Data to Augment Visualization Recommendations}}

Although current literature does not cover the entire visualization design space, we can still apply existing \revised{graphical perception guidelines} to visualization recommendation systems.
Here we use three representative visualization recommendation systems as examples to demonstrate how our survey data can inform encoding decisions.

\subsubsection{Draco~\cite{Moritz2018formalizing}}
\label{subsec:inform-draco}

The Draco system models visualization design guidelines as hard or soft constraints.
Draco first excludes the visualizations that violate the hard constraints and then searches for the most preferred visualizations using soft constraints.
Although Draco already applies some of the existing visualization design knowledge in its applications (Draco-APT~\cite{Mackinlay1986automating}, Draco-CQL~\cite{Mackinlay1986automating,Mackinlay2007showme}, and Draco-Learn~\cite{Kim2018assessing,Saket2018task}), the number of utilized research papers is limited.
In this paper, we collect \revised{graphical} perceptual results from 59 existing literature works.
\revised{First}, our synthesized guidelines in~\autoref{sec:literature-review} could be translated into hard or soft constraints to further enhance Draco's recommendations.
For example, we translate two visualization design rules from \autoref{subsec:within-chart} (e.g., preferring color encodings for the \texttt{aggregate} task) into Draco soft constraints in 
\autoref{draco-exp}.


\begin{lstlisting}[language=asp,style=base,caption={Examples of translating our synthesized guidelines into Draco soft constraints~\cite{Moritz2018formalizing}.},label=draco-exp]
*%Prefer to use fewer encodings with fields*
~soft~(encoding_field,E) :- encoding(E), field(E,_).
*%Prefer to use color for aggregate task*
~soft~(aggregate_color,E) :- task(aggregate), channel(E,color).
\end{lstlisting}

\revised{Second, our synthesized dataset of perceptual results (in~\autoref{sec:method:schema}) can be used as an input corpus for Draco-Learn. 
We provide scripts in the supplemental material to automatically translate our datasets into pairs of ranked visualizations as the training dataset for Draco-Learn.
By learning visualization rankings from a large number of research papers (instead of two papers), Draco-Learn could potentially support more chart types and produce more effective recommendations for observed data characteristics and task types.}

\begin{figure}
  \centering
    \begin{subfigure}[b]{0.49\columnwidth}
    \centering
    \includegraphics[width=\textwidth]{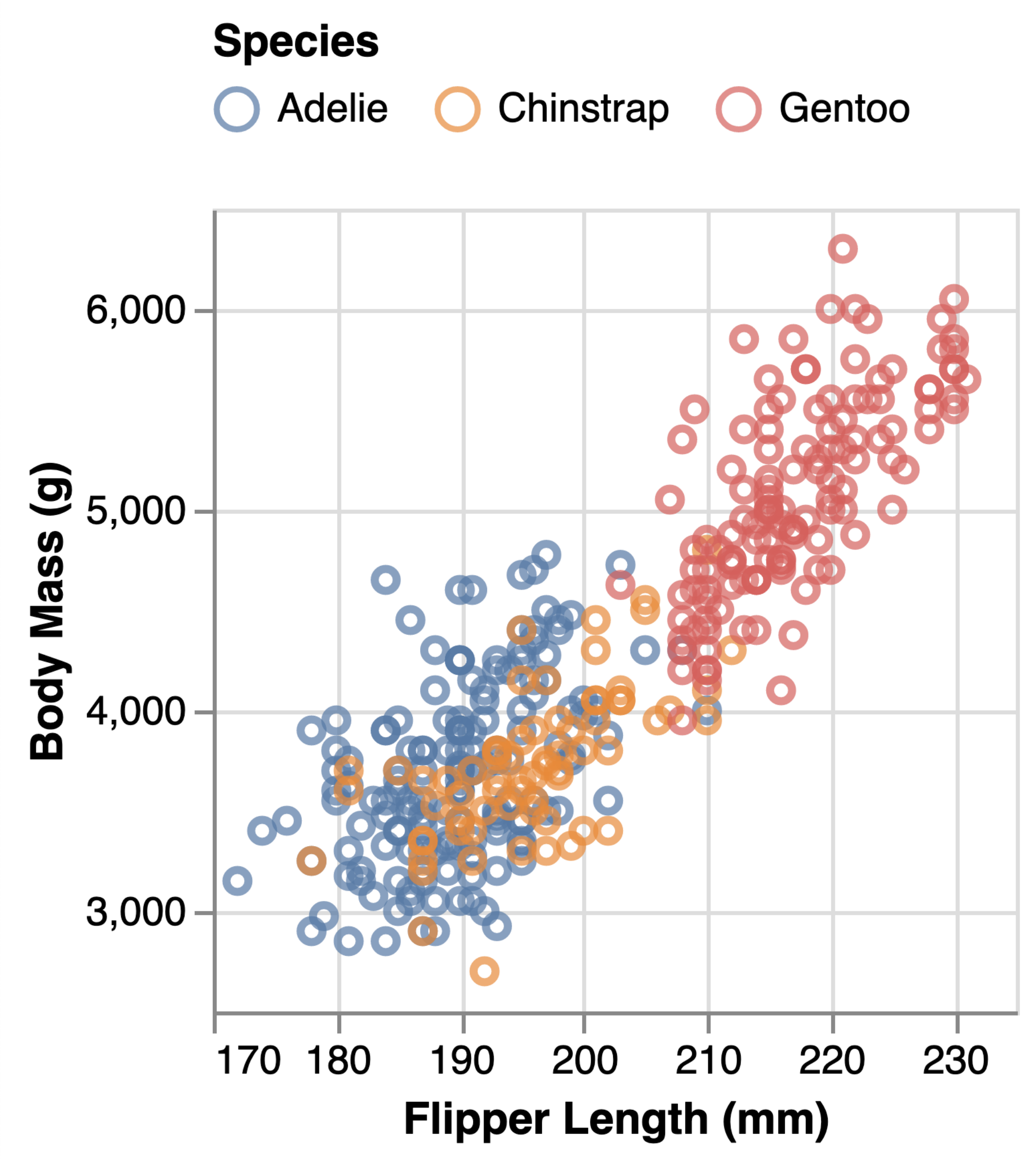}
    \caption{Voyager Recommendation}
    \label{fig:voyager-original}
    \end{subfigure}
    \begin{subfigure}[b]{0.49\columnwidth}
    \centering
    \includegraphics[width=\textwidth]{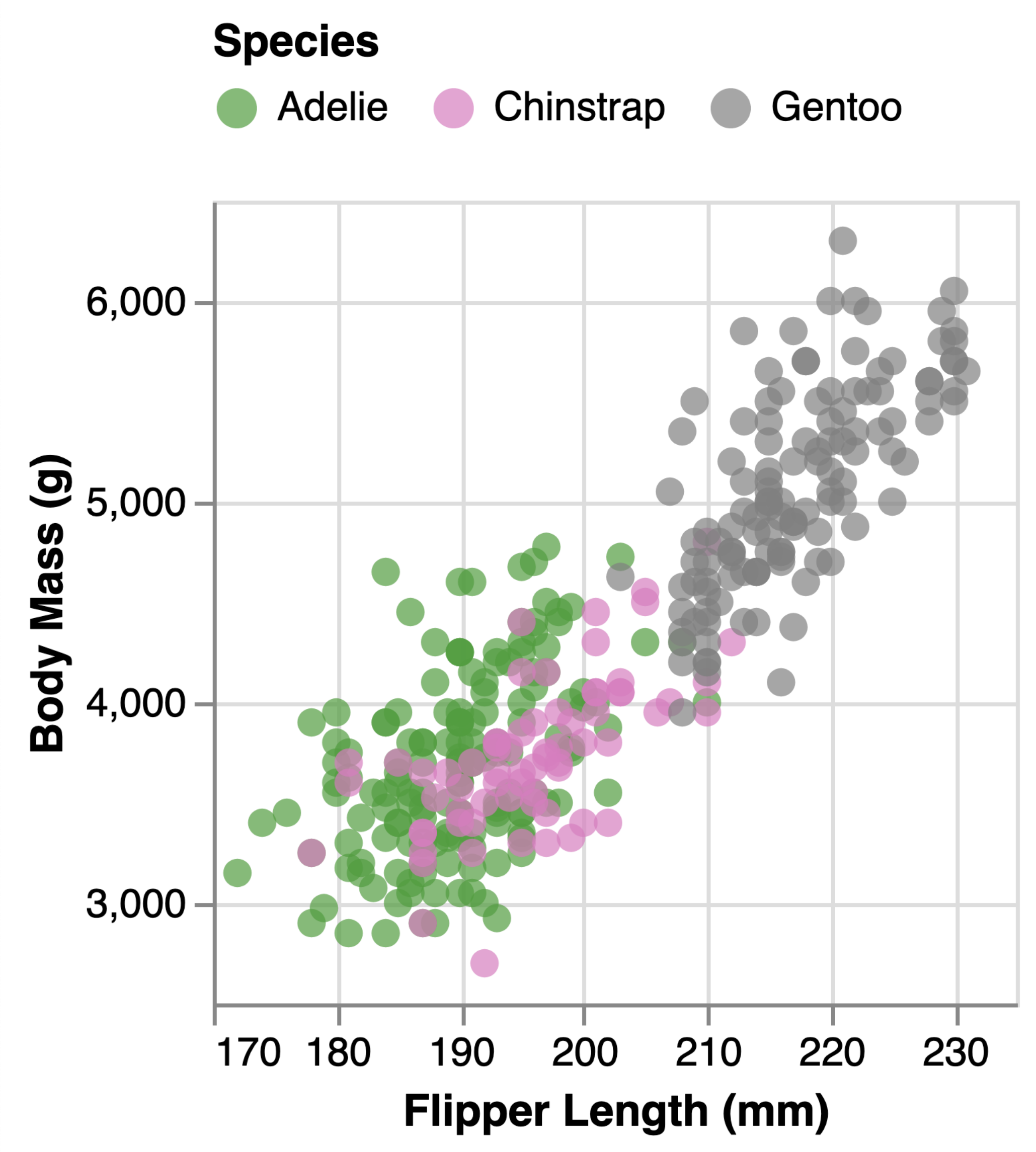}
    \caption{Improved Recommendation}
    \label{fig:voyager-improved}
    \end{subfigure}
  \caption{An example of improving Voyager's~\cite{Wongsuphasawat2015voyager,Wongsuphasawat2017voyager2} recommendations using our synthesized guidelines (specifically, using \cite{Demiralp2014learning,Smart2019measuring}).}
  \label{fig:voyager}
  \Description{Figure 5 shows an example of how to improve the visualizations recommended by Voyager systems using our synthesized guidelines.}
\end{figure}

\subsubsection{Voyager~\cite{Wongsuphasawat2015voyager,Wongsuphasawat2017voyager2}}
\label{subsec:inform-voyager}

The Voyager system suggests both data attributes and visual encodings.
Voyager uses CompassQL~\cite{Wongsuphasawat2016towards} as the recommendation engine and Vega-lite~\cite{Satyanarayan2017vegalite} as the visualization renderer.
\autoref{fig:voyager-original} shows one of the visualizations recommended by Voyager, which is a colored scatterplot using a Vega-lite color palette (\scalerel*{\includegraphics{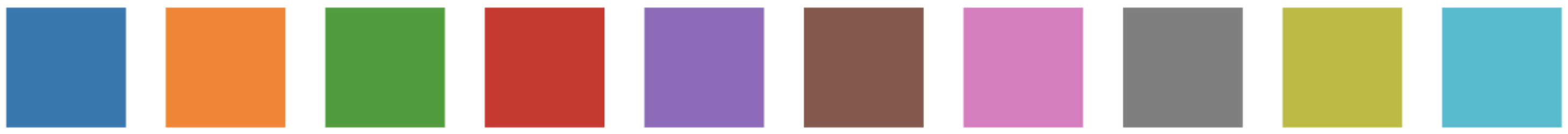}}{B}) for nominal data.
We re-generate the recommended chart (as shown in~\autoref{fig:voyager-improved}) using the design guidelines from~\autoref{subsec:encoding-outcomes}: (1) the re-ordered color palette (\scalerel*{\includegraphics{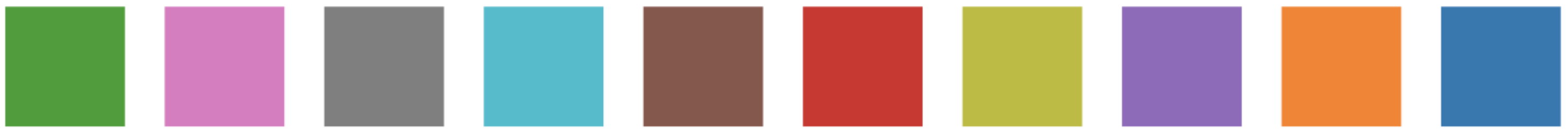}}{B}) can maximize perceptual distance~\cite{Demiralp2014learning}; and (2) colors are more discriminable with filled shapes~\cite{Smart2019measuring}.
We can see that each category (species) is more distinguishable from the other in the improved recommendation compared to the Voyager original recommendation (\autoref{fig:voyager}).
In the same spirit, we can use suggested palettes for \textsl{shape}, \textsl{color hue}, and \textsl{area} from existing literature~\cite{Demiralp2014learning,Gramazio2016colorgorical,Fang2016categorical,Wang2018optimizing} to improve the effectiveness of Voyager's recommended visualizations.
\revised{We provide the Vega-Lite~\cite{Satyanarayan2017vegalite} specifications for suggested color palettes in the supplemental material. They can be ingested by visualization recommendation systems that use Vega-Lite as a visualization renderer, similar to existing work (e.g., \cite{Wongsuphasawat2015voyager,Wongsuphasawat2017voyager2,Lin2020dziban,Moritz2018formalizing,Kim2017graphscape}).}

\begin{figure}[H]
\centering
 \includegraphics[width=1.0\columnwidth]{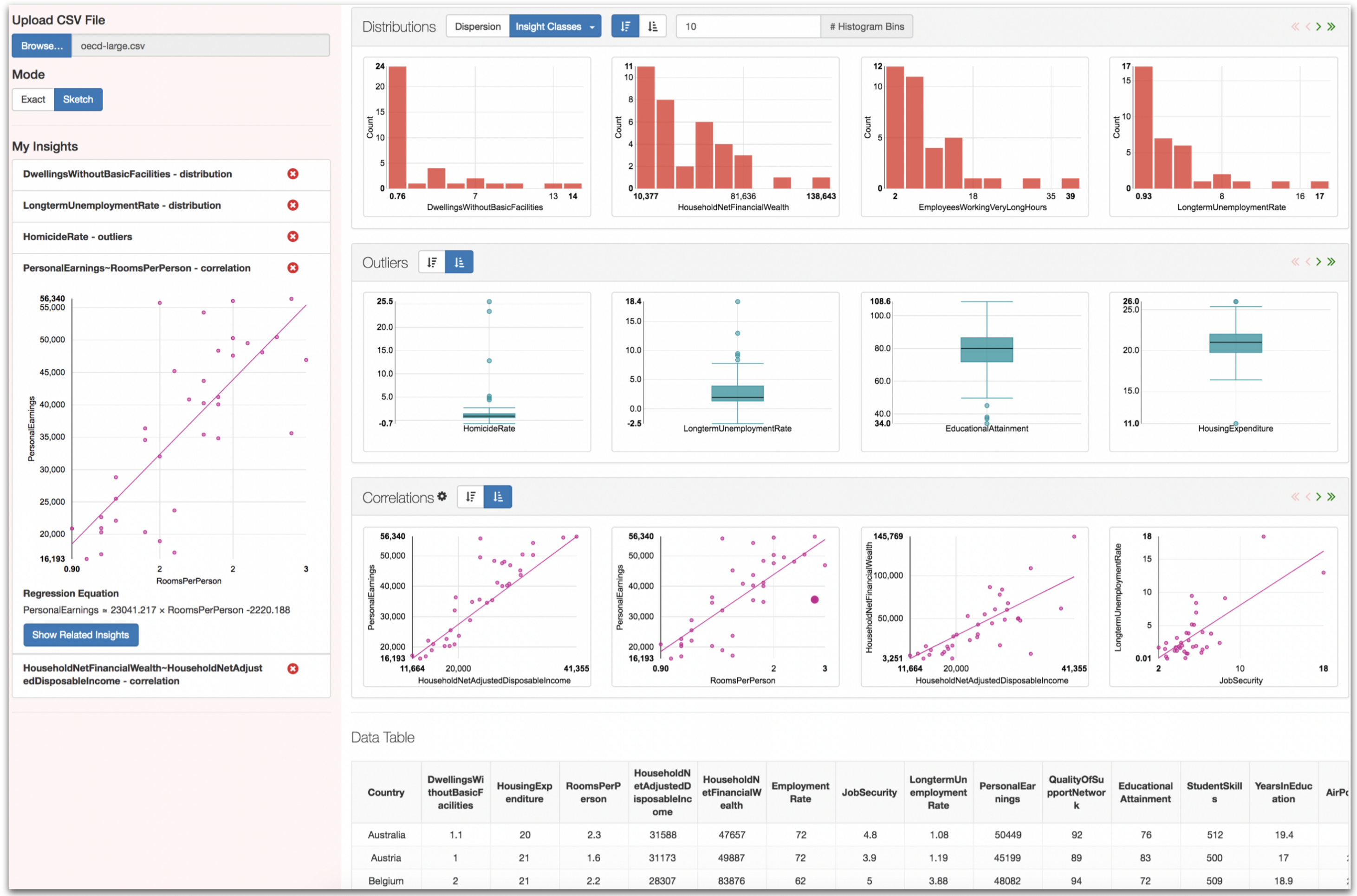}
 \caption{A screenshot of the Foresight system~\cite{Demiralp2017foresight} showing 3 of the 12 supported insight classes: heavy tails, outliers, and correlations. }
 \label{fig:foresight}
 \Description{Figure 6 shows a screenshot of the Foresight system. They suggest different visualizations according to different data characteristics.}
\end{figure}

\subsubsection{Foresight~\cite{Demiralp2017foresight}}
\label{subsec:inform-foresight}

The Foresight system suggests data attributes based on their ``insight'' scores and presents the recommendation with either a bar chart, a box plot, or a scatterplot.
As we can see from~\autoref{fig:foresight}, Foresight uses different chart types to visualize different ``insight classes''.
Bar charts are used for distributions (heavy tails), while box plots are used for outliers and scatterplots for correlations.
However, the results from experiments by Saket et al.~\cite{Saket2018task} suggest that scatterplots perform the best for \texttt{finding anomalies}, while line charts for \texttt{correlation}, and bar charts for \texttt{distribution} tasks.
Thus, replacing existing chart types with the best ones (see~\autoref{tab:between-charts-outcomes}) based on empirical results for corresponding tasks might help users gain more insights more effectively with Foresight.

\subsection{Challenges \& Open Research Areas}

Here we \revised{synthesize our takeaways from \autoref{sec:literature-review} into open research} challenges toward a comprehensive ranking of visualization designs and discuss directions for future research based on our analysis.

\subsubsection{Gaps in Visualization Comparison Coverage}

\paragraph{Gaps in theoretical work.}
As mentioned in \autoref{sec:method:schema}, theory work is critical to generalizing the findings of specific experiments as the takeaways can be applied broadly in visualization.
Although all twelve encodings are ranked (or pruned) according to theoretical hypotheses (see~\autoref{tab:encoding-coverage},~\autoref{tab:mackinlay-encodings}), only a few visualization types (scatterplot, bar chart, line chart, and cartogram) are discussed in theory work (see~\autoref{tab:chart-coverage}).
Other chart types, such as area charts, pie charts, and heatmaps, are never ranked theoretically.
In other words, interference effects among visual encodings are rarely theoretically studied.
\textbf{\texttt{Solutions:}} An impactful area would be deriving theoretical principles from existing experiment results for multi-encoding designs to infer how well different encodings will work \textit{together} and whether the performance of encoding \textit{combinations} still follow the same rankings. New theories can also help to prune the design space to identify gaps that truly warrant new experiments.

\paragraph{Gaps in experimental work.}
On the one hand, we observe much fewer experiments evaluating how effectively each encoding could convey ordinal data (only CH and CS are tested).
\textbf{\texttt{Solutions:}} We urge more empirical work to conclude which encoding to pick under different task scenarios.

On the other hand, existing evaluations mainly focus on specific encodings (PX, PY, L, Ar, CS, CH) or charts (scatterplots and bar charts), while other chart types are either only compared with one or two other charts, or never studied with different variants (see~\autoref{sec:litrev-charts}).
\textbf{\texttt{Solutions:}} Evaluating the performance of previously ignored encodings (e.g., T, S) and chart types (e.g., area charts, heatmap) under different analysis tasks would contribute more ``ground truth'' evidence to further validate our approaches to automating the visualization design process.

\subsubsection{Inconsistencies and Conflicts in the Literature}

\paragraph{Between theories and experiments.}
We observe that theoretical hypotheses might not necessarily be ``correct'' in a practical sense.
For example, as previously mentioned, five attribute-encoding pairs ((Q, T/S), (O, S), (N, Ar/CS) are considered inexpressive based on Mackinlay's work~\cite{Mackinlay1986automating} (see~\autoref{tab:mackinlay-encodings}); however, a more recent evaluation~\cite{Chung2016how} shows different results.
Mackinlay suggests that T and S are not relevant to quantitative data, but according to the results from Chung et al.'s experiments, both T and S encodings perform better in \textit{accuracy} than O conveying quantitative data with \texttt{estimate trend} and \texttt{find extremum} tasks.
\textbf{\texttt{Solutions:}} Refining core theory work in light of recent experimental results could further enhance the performance of visualization recommendation systems.

\paragraph{Between different experiments.}
Even when experiments were similar, we may find contradictory results.
Even though Godau et al.~\cite{Godau2016perception} and Xiong et al.~\cite{Xiong2019biased} both conducted experiments to test human bias in perceiving average position for length (bar charts) and position encodings (scatterplot or line charts), they have completely different results (as shown in~\autoref{tab:bias}).
Godau et al. only find underestimation in bar charts but no bias for point positions (scatterplots).
However, Xiong et al. find significant bias in both bar charts and line charts, where line positions are underestimated while bar positions are overestimated.
In another example, Harrison et al.~\cite{Harrison2014ranking} find Weber's law to be a convincing model for how people perceive data correlations; however, in a re-analysis of the same data, Kay and Heer~\cite{Kay2015beyond} find Weber's Law not to be a good fit. It is natural in science to improve upon existing results and theories; however, there is currently no easy way to identify and track these discrepancies within the literature and translate them into concrete improvements to visualization recommendation systems.
\textbf{\texttt{Solutions:}} Redesigning experiments to test visualizations with controversial results, conducting more comprehensive comparisons between more nuanced design decisions, and involving more metrics could lead to more precise visualization design rankings for recommendation systems.

\paragraph{\textbf{Summary.}}
Given the (multiple) discrepancies we have observed, we argue that the findings of both visualization theory and experimental research should be treated as \textit{hypotheses} until subsequent experiments converge on a consistent set of results. Furthermore, we argue that replication experiments should be held in high regard within the visualization community regardless of whether their findings reinforce or challenge our current assumptions, since either way, they are the \textit{only} way to validate our understanding of how people perceive and use visualizations. We need them to ensure that visualization recommendation algorithms are built upon a solid foundation of theoretical and empirical findings; we should reward them accordingly.

\subsection{Limitations \& Future Work}

Our literature review contributes a detailed record of how different visualization designs are compared and ranked in 59 different theoretical and experimental papers. 
This record not only specifies all researched visualization designs but also keeps track of the ranking of their performance (\textit{accuracy}, \textit{bias}, \textit{JND}, \textit{time}, \textit{user-preference}) under different task scenarios.
A next step to extend this work could be to apply the findings to develop better encoding strategies within visualization recommendation systems, such as adapting the recommendation strategy based on the user's current analysis task.

Given our initial goal is to understand how different visualization designs (especially different encodings) are ranked in the current theoretical and experimental work, our schema only records \{\textit{data types}, \textit{data characteristics}, \textit{data transformations}, \textit{encoding channels}, \textit{mark types}, \textit{scales}\} for each covered design (details in \autoref{lst-covered-design}).
For more granular design decisions, we add notes to specify them.
For example, to record Talbot et al.'s experiments testing bar charts~\cite{Talbot2014four}, 
we add notes to specify each variant of the bar chart tested, such as whether two bars are aligned or separated, whether distractors are added, the indicator location, etc.
However, it is hard to parse these notes automatically.
We see our schema as a starting point for collating existing encoding design knowledge and encourage the visualization community to extend this schema to support more nuanced visualization designs.

Informed by existing work on visualization design spaces and \revised{graphical} perception studies, we excluded (1) 3D visualizations, (2) graph visualizations, and (3) visualizations with animations or interactions from our defined visualization space.
When more theoretical and experimental findings become available in the literature, expanding our work to include these excluded designs would be interesting.

We also note that by focusing on \revised{graphical} perception, we are unable to account for other factors that may influence the overall effectiveness of a visualization design, such as visual aesthetics~\cite{Tufte2001visual}, intuition, and metaphors~\cite{Ziemkiewicz2008shaping}, as well as user background and preferences~\cite{Ziemkiewicz2012understanding}. Developing a broader framework encompassing both \revised{graphical} perception and these other factors would be exciting future work.

\begin{acks}
The authors wish to thank colleagues in the UMD HCIL and the BAD Lab, as well as Niklas Elmqvist, Michael Correll, Steve Franconeri and our paper reviewers for their thoughtful feedback. This work was supported in part by NSF award IIS-1850115 and a VMWare Early Career Faculty Grant.
\end{acks}

\bibliographystyle{ACM-Reference-Format}
\bibliography{reference}

\end{document}